\documentclass[aps,prd,amsmath,amssymb,twocolumn,preprintnumbers,nofootinbib]{revtex4-1}

\usepackage[utf8]{inputenc} 
\usepackage{graphicx}
\usepackage{multirow}
\graphicspath{{./figures/}}
\usepackage{url}
\usepackage[bookmarks, pagebackref=false]{hyperref}
\usepackage[usenames,dvipsnames]{xcolor}
\definecolor{orange}{cmyk}{0,0.5,1,0}
\definecolor{rossoCP3}{cmyk}{0,.88,.77,.40}
\definecolor{graa}{rgb}{0.8,0.8,0.8}
\definecolor{blaa}{rgb}{0.2,0.2,0.6}
		\hypersetup{
			colorlinks,
			bookmarksopen,
			bookmarksnumbered,
			citecolor=blaa, 		
			linkcolor=rossoCP3,	
			urlcolor=rossoCP3,			
			}
\usepackage{amsthm}
\usepackage{bm}
\usepackage{bbm}
\usepackage{pxfonts}

\usepackage{amsmath,amssymb,amsfonts}
\usepackage{color}
\usepackage{float}
\usepackage{hyperref}
\usepackage[Symbolsmallscale]{upgreek}
\usepackage{amsmath}
\usepackage{amsfonts}
\usepackage{amssymb,dsfont}
\usepackage{graphicx}
\usepackage{amssymb}
\usepackage[vcentermath]{youngtab}
\usepackage[all]{xy}
\usepackage{dsfont}%
\usepackage{placeins}
\usepackage{xspace}
\usepackage{cancel} 

\usepackage{slashed}
\usepackage[caption=false, labelformat=simple, listofformat=subsimple, labelfont=default, margin=5pt, justification=raggedright]{subfig}

	\makeatletter
		\renewcommand{\p@subfigure}{}
	\makeatother

\usepackage{natbib}

\newcommand{\MSb}{\overline{\textrm{MS}}}

\newcommand{\gGF}{g_{\rm GF}^2}
\newcommand{\gGFl}[1]{g_{{\rm GF},#1}^2}

\newcommand{\beq}{\begin{eqnarray}}
\newcommand{\eeq}{\end{eqnarray}}

\newcommand{\bmp}{\noindent\begin{minipage}{16cm}}
\newcommand{\emp}{\end{minipage}\vskip 7mm} 

\def\lsim{\mathrel{\rlap{\lower4pt\hbox{\hskip1pt$\sim$}}
    \raise1pt\hbox{$<$}}}                
\def\gsim{\mathrel{\rlap{\lower4pt\hbox{\hskip1pt$\sim$}}
    \raise1pt\hbox{$>$}}}                

\newcommand{\of}[1]{\left(#1\right)}

\newcommand{\bof}[1]{\biggl(\bigg.#1\bigg.\biggr)}

\newcommand{\sof}[1]{\bigl(\big.#1\big.\bigr)}
\newcommand{\ssof}[1]{(#1)}

\newcommand{\ssfof}[1]{[#1]}
\newcommand{\cof}[1]{\left\{#1\right\}}

\newcommand{\avof}[1]{\left\langle #1\right\rangle}

\newcommand{\tsqrt}[1]{\smash{\sqrt{#1}}}
\renewcommand*\[{\begin{equation}}
\renewcommand*\]{\end{equation}}
\newcommand{\order}{\mathcal{O}}

\definecolor{newgreen}{RGB}{10,100,20}

\begin{document}


\title{\texorpdfstring{\Large\color{rossoCP3}    Safety versus triviality on the lattice}{ }}

\author{Viljami Leino}
\email{viljami.leino@tum.de}
\affiliation{Physik Department, Technische Universit\"at M\"unchen, 85748 Garching, Germany}
\author{Tobias Rindlisbacher}
\email{tobias.rindlisbacher@helsinki.fi}
\affiliation{Department of Physics \& Helsinki Institute of Physics,
P.O. Box 64, FI-00014 University of Helsinki}
\author{Kari Rummukainen}
\email{kari.rummukainen@helsinki.fi}
\affiliation{Department of Physics \& Helsinki Institute of Physics,
P.O. Box 64, FI-00014 University of Helsinki}
\author{Francesco Sannino}
\email{sannino@cp3.dias.sdu.dk}
\affiliation{CP$^{ \bf 3}$-Origins \&  Danish IAS,  University of Southern Denmark }
\author{Kimmo Tuominen}\email{kimmo.i.tuominen@helsinki.fi}
\affiliation{Department of Physics \& Helsinki Institute of Physics,
P.O. Box 64, FI-00014 University of Helsinki}

\begin{abstract}
We present the first numerical study of the ultraviolet dynamics of non-asymptotically free gauge-fermion theories at large number of matter fields. As testbed theories we consider non-abelian SU(2) gauge theories with 24 and 48 Dirac fermions on the lattice.
For these number of flavors asymptotic freedom is lost and the theories
are governed by a gaussian fixed point at low energies. In the ultraviolet they can develop a physical cutoff and therefore be trivial, or achieve an interacting safe fixed point and therefore be fundamental at all energy scales.
We demonstrate that the gradient flow method can be successfully implemented and applied to determine the renormalized running coupling
when asymptotic freedom is lost. Additionally, we prove that our analysis is
connected to the gaussian fixed point as our results nicely
match with the perturbative beta function. Intriguingly, we observe that it
is hard to achieve large values of the renormalized coupling on the lattice.
This might be an early sign of the existence of a physical cutoff and
imply that a larger number of flavors is needed to achieve the safe fixed point. A more conservative interpretation of the results is that the current lattice action is unable to explore the deep ultraviolet region where safety might emerge. Our work constitutes an essential step towards determining the ultraviolet fate of non asymptotically free gauge theories.
\vskip .1cm
{\footnotesize  \it Preprint:  HIP-2019-22/TH, TUM-EFT 127/19 and CP3-Origins-2019-29 DNRF90
}
\end{abstract}
\maketitle

\section{Introduction}

Asymptotically free theories \cite{Gross:1973ju,Politzer:1973fx} are fundamental according to Wilson~\cite{Wilson:1971bg,Wilson:1971dh} since they are well defined from low to arbitrary high energies. This remarkable property stems simply from the fact that asymptotically free theories are obviously conformal (and therefore scale invariant) at short distances, given that all interaction strengths vanish in that limit. This is one of the reasons why asymptotic freedom has played such an important role when building extensions of the Standard Model (SM). Non-abelian gauge-fermion theories, like QCD, with a sufficiently low number of matter fields are time-honored examples of asymptotically free gauge theories. These theories feature a single four-dimensional marginal coupling induced by the gauge dynamics and no further interactions are needed to render the theories asymptotically free.

On the other hand, purely scalar and scalar-fermion theories are not asymptotically free. Adding elementary scalars, and upgrading gauge-fermion theories to gauge-Yukawa systems, one discovers that scalars render the existence of asymptotic freedom less guaranteed. In particular, complete asymptotic freedom in all marginal couplings is no longer automatically ensured by requiring a sufficiently low number of scalar and fermion matter fields.

To determine the asymptotically free conditions on the low energy values of the accidental couplings that may lead to complete asymptotic freedom a one-loop analysis in all marginal couplings is sufficient.
One discovers that the gauge-interactions are essential to tame the unruly behavior of the accidental couplings provided the latter start running within a specific region in coupling space at low energies.

Non-asymptotically free theories can belong either to the trivial or the safe category. Triviality occurs when the theories develop a physical cutoff and can therefore be viewed as low energy effective descriptions of a more fundamental but typically unknown quantum field theory. Triviality literally means that the only way to make sense of these theories as fundamental theories (when trying to remove the cutoff) is by turning off the interactions. In truth it is rather difficult to demonstrate that a theory is trivial beyond perturbation theory since the couplings become large in the UV and therefore one can imagine a non-perturbative UV fixed point to emerge leading to non-perturbative asymptotic safety.
Nevertheless, we can be confident that certain theories are indeed trivial. Perhaps the best known example is the 4-dimensional $\lambda \phi^4$ theory, which was studied on the lattice in a series of papers by L\"uscher and Weisz \cite{Luscher:1987ay,Luscher:1987ek, Luscher:1988uq}.  The triviality here was established using large order hopping parameter expansion with perturbative renormalization group evolution, and corraborated with several lattice Monte Carlo simulations (see, for example, \cite{Montvay:1988uh,Wolff:2009ke}).  
Related to our study here, the triviality versus conformality has also been investigated in large-$N_f$ QCD with staggered fermions \cite{deForcrand:2012vh}.

Analytically, using a-maximisation and violation of the a-theorem, one can also demonstrate that certain non-asymptotically free  supersymmetric gauge-Yukawa theories such as super QCD with(out)
a meson and their generalisations are trivial once asymptotic freedom is lost by adding enough super matter fields \cite{Intriligator:2015xxa}.

We move now to safe theories. These achieve UV conformality while remaining interacting meaning that the interaction strengths freeze in the UV without vanishing. The first four dimensional safe field theories were discovered in \cite{Litim:2014uca} within a perturbative study of gauge-Yukawa theories in the Veneziano limit of large number of flavors and colors. This discovery opened the door to new ways to generalise the Standard Model as envisioned first in \cite{Sannino:2015sel} and then investigated in \cite{Abel:2017ujy,Abel:2017rwl,Pelaggi:2017wzr,Mann:2017wzh,Pelaggi:2017abg,Bond:2017wut,Abel:2018fls} with impact in dark matter physics \cite{Sannino:2014lxa,Cai:2019mtu} and cosmology \cite{Pelaggi:2017abg}. Additionally it allowed to use the large charge \cite{Hellerman:2015nra,Alvarez-Gaume:2016vff} method to unveil new controllable CFT properties for four-dimensional non supersymmetric quantum field theories \cite{Orlando:2019hte}.

Although both safe and free theories share the common feature of having no cutoff, the respective mechanisms and dynamics for becoming fundamental field theories are dramatically different \cite{Litim:2014uca}. For example, within perturbation theory it is impossible to achieve safety with gauge-fermion theories \cite{Caswell:1974gg}. Yukawa interactions and the consequent need for elementary scalars is an essential ingredient to tame the UV behaviour of these theories. Of course, this is a welcome discovery given that it provides a pleasing theoretical justification for the existence of elementary scalars, such as the Higgs, and Yukawa interactions without the need to introduce baroque symmetries such as supersymmetry. Beyond perturbation theory, however, little is known and it is therefore worth asking whether scalars are needed to achieve asymptotic safety.

Interesting hints come from the knowledge of the beta function for abelian and non-abelian gauge-fermion theories at leading order in
$1/N_f$ \cite{PalanquesMestre:1983zy,Gracey:1996he,Holdom:2010qs,Pica:2010xq}. This result suggests, as we shall review below, the potential existence of short distance conformality \cite{Pica:2010xq,Antipin:2017ebo}. However due to the fact that the UV zero in the beta function stems from a logarithmic singularity and more generally that the beta function is not a physical quantity, careful consistency checks of these results are crucial~\cite{Ryttov:2019aux}.

For these reasons, and because it is important to uncover the phase diagram of four dimensional quantum field theories, we initiate here a consistent lattice investigation of the ultraviolet fate of non-abelian gauge-fermion theories at a small number of colours but at large number of flavours where asymptotic freedom is lost. These parameters are currently inaccessible with other methods. Specifically, we will investigate the SU(2) gauge theory with $24$ and $48$ massless Dirac flavours. These two numerical values are substantially larger than the value where asymptotic freedom is lost which is $11$, and they are also very roughly estimated to be close to the region where the $1/N_f$ squared corrections become relevant \cite{Antipin:2017ebo,Holdom:2010qs}. This was taken to be a sign of where one would expect the UV fixed point to disappear \cite{Antipin:2017ebo} and the theory develop a cutoff.

To numerically investigate the UV properties of these two theories, in our
lattice analysis we focus on the determination of the renormalized running gauge coupling. This is achieved by implementing the Yang-Mills gradient flow method at finite volume and with Dirichlet boundary conditions \cite{Ramos:2015dla}, enabling us to perform lattice simulations at vanishing fermion mass.  We are able to demonstrate that our simulations are performed
in the physical region connected to the gaussian fixed point. This is
corroborated by the fact that our results match onto the scheme-independent
two-loop perturbative beta function in the infrared. We also discover that it is difficult to achieve large values of the renormalized coupling on the lattice. This might be interpreted as an early sign of the existence of a physical cutoff both for 24 and 48 flavors and that a larger number of flavors would be needed to achieve the safe fixed point. However, a closer look at the large $N_f$ predicted UV behaviour of these theories, if applicable, would suggest a safe fixed point to occur at a much stronger value of the gauge coupling\footnote{We observe that the leading $N_f$ beta function is scheme-independent.} than achievable with the current simulations. This suggests a more conservative interpretation, that we are not yet able to explore the deep ultraviolet region where safety might emerge. Nevertheless we feel that our work constitutes a necessary stepping stone towards unveiling the ultraviolet fate of non-asymptotically free gauge theories.

The paper is organised as follows: In Section \ref{CW} we briefly review the analytical results for (non)-abelian gauge theories at leading order in $1/N_f$. We compare the large $N_f$ beta function with two-loop perturbation theory and discuss the conformal window 2.0 \cite{Antipin:2017ebo}. It is straightforward to specialise the results of this section to the case of SU(2) gauge theory investigated on the lattice. Section \ref{SoL} is constituted by several subsections with the goal to make it easier for the reader to focus on the relevant aspects of the lattice setup and results: We begin with an introduction to the lattice action its features. We then discuss how the gauge coupling and its running is defined through the Yang-Mills gradient flow method. Then we summarise the lattice results for $N_f = 24$ and $N_f = 48$. In Sec.~\ref{Conc} we offer our conclusions and directions for further studies.


\section{Conformal Window 2.0: review of the analytic results}\label{CW}

Consider an $SU(N_c)$ gauge theory with $N_f$ fermions transforming according to a given representation of the gauge group. Assume that asymptotic freedom is lost, meaning that the number of flavours is larger than $N_f^{AF} = 11C_G/(4T_R)$, where the first coefficient of the beta function changes sign. In the fundamental representation the relevant group theory coefficients  are  $C_G=N_c$, $C_R=(N_c^2-1)/2N_c$ and $T_R=1/2$. At the one loop order the theory is free in the infrared, i.e. non-abelian QED, and simultaneously trivial. As discussed in the introduction, this means that the theory has a sensible continuum limit, by sending the Landau pole induced cutoff to infinity, only if the theory becomes non-interacting at all energy scales.

At two-loops, in a pioneering work, Caswell~\cite{Caswell:1974gg} demonstrated that the sign of the second coefficient of the gauge beta function is such that an UV interacting fixed point, which would imply asymptotic safety, cannot arise when the number of flavours is just above the value for which asymptotic freedom is lost.  This observation immediately implies that for gauge-fermion theories triviality can be replaced by safety only above a new critical number of flavours.  In order to investigate this possibility, consider the large  $N_f$-limit at fixed number of colours. The leading order large $N_f$ beta functions for QED and non-abelian gauge theories were constructed in~\cite{PalanquesMestre:1983zy,Gracey:1996he}, while a summary of the results and
possible investigation for the existence of UV fixed points appeared first in~\cite{Holdom:2010qs,Pica:2010xq} with the scaling exponents computed first in~\cite{Litim:2014uca}. Although in this work we will concentrate on non-abelian gauge theories we now briefly comment on the status of QED. Even though the large $N_f$ beta function develops a non-trivial zero, it was demonstrated in Ref.~\cite{Antipin:2017ebo}, that at the alleged UV fixed point the fermion mass anomalous dimension violates the unitarity bound and hence the UV fixed point is unphysical. At this order in $1/N_f$  we conclude that QED is   trivial.

For the non-abelian case, using the conventions of~\cite{PalanquesMestre:1983zy,Holdom:2010qs}, the standard beta function reads
 \begin{equation}
\beta(\alpha)\equiv\frac{\partial \ln \alpha}{\partial \ln \mu}=-b_1\frac{\alpha}{\pi}+...,  \qquad \alpha= \frac{g^2}{4\pi} \ ,
\label{e10}\end{equation}
with $g$ the gauge coupling. At large $N_f$ it is convenient to work
in terms of the normalised coupling   $A\equiv N_f T_R\alpha/\pi$. Expanding in  $1/N_f$ we can write
\begin{equation}
\frac{3}{2}\frac{\beta(A)}{A}=1+\sum_{i=1}^{\infty}\frac{H_i(A)}{N_f^i} \ ,
\label{e8}\end{equation}
where the identity term corresponds to the one loop result and constitutes the zeroth order term in the $1/N_f$ expansion. If the functions $|H_i(A)|$ were finite, then in the large $N_f$ limit the zeroth order term would prevail and the Landau pole would be inevitable. This, however, is not the case due to the occurrence of a divergences in the $H_i(A)$ functions.

According to the large $N_f$ limit, each function $H_i(A)$ re-sums an infinite set of Feynman diagrams at the same order in $N_f$ with $A$ kept fixed. To make this point explicit, consider the leading $H_1(A)$ term. The $n{\rm{th}}-$loop beta function coefficients $b_n$ for $n \geq 2$ are polynomials of order $n-1$ in $T_R N_f$:
\begin{equation}
b_n=\sum_{k=0}^{n-1} b_{n,k} (T_R N_f)^k \ .
\label{pert}
\end{equation}
The coefficient with the highest power of $T_R N_f$ will be $b_{n,n-1}$ and this is the coefficient contributing to $H_1(A)$ at the $n{\rm{th}}-$loop order. Moreover,
it was shown in ~\cite{Shrock:2013cca} that the $b_{n,n-1}$ terms are invariant under the scheme transformations that are independent of $N_f$ (as appropriate for large-$N_f$ limit).

Now, the $n{\rm{th}}-$loop beta function will have an interacting UV fixed point (UVFP) when the following  equation has a physical zero~\cite{Pica:2010xq}
\begin{equation}
b_1 + \sum_{k=2}^{n} b_k \alpha^{k-1} = 0 \  \ \ \text{where} \ \ b_1=\frac{\beta_0}{2}=\frac{11C_G}{6}-\frac{2T_R N_f}{3} \ .
\label{BF}
\end{equation}
This expression simplifies at large $N_f$. Truncating at a given perturbative order $n_{\rm{max}}$ one finds that the highest loop beta function coefficient $b_{n_{\rm{max}}}$ contains just the highest power of $(T_R N_f)^{n_{\rm{max}}-1}$ multiplied by the coefficient
$b_{n_{\rm{max}}, n_{\rm{max}}-1}$, as can be seen from Eq.\eqref{pert}. Since this highest power of $(T_R N_f)^{n_{\rm{max}}-1}$ dominates in the $N_f\to \infty$ limit and since in this limit $b_1<0$, the criterium for the existence of a UV zero in the $n_{\rm{max}}-$loop beta function
becomes~\cite{Pica:2010xq}:
\emph{\begin{center} for $N_f\to \infty \ , \ \  $ $\beta(\alpha)$ has an UVFP only  if $b_{n_{\rm{max}}, n_{\rm{max}}-1}>0$ \ .\end{center}}

In perturbation theory, only the first few coefficients $b_{n,n-1}$ are known but, remarkably, it is possible to resum the perturbative infinite sum to obtain $H_1(A)$. From the results in~\cite{PalanquesMestre:1983zy,Gracey:1996he}
\begin{eqnarray}
H_1(A)&=&-\frac{11}{4}\frac{C_G}{T_R}+\int_0^{A/3} I_1(x)I_2(x)dx,\label{e7}\\
 I_1(x)&=&{\frac {  ( 1+x ) ( 2\,x-1 )^2  ( 2\,x-3 )^2 \sin ( \pi \,x )^{3} \Gamma
 ( x-1 )^{2}\Gamma  ( -2\,x ) }{ ( x-2 ) \ {\pi }^{3}}} \nonumber \\
I_2(x)&=&\frac{C_{{R}}}{T_R}+\frac{\left( 20-43\,x+32\,{x}^{2}-14\,{x}^
{3}+4\,{x}^{4} \right) }{ 4\left( 2\,x-1 \right)  \left( 2\,x-3 \right) \left( 1-x^2
 \right) }\frac{C_{{G}}}{T_R}  \ .\nonumber
\end{eqnarray}
By inspecting  $I_1(x)$ and $I_2(x)$ one notices that the  $C_G$ term in $I_2$ has a pole in the integrand at $x=1$ ($A=3$). This corresponds to a logarithmic singularity in $H_1(A)$, and will cause the beta function to have a UV zero already at this order in the $1/N_f$ expansion and, by solving $1+H_1(A)/N_f=0$, this non-trivial UV fixed point occurs
at~\cite{Litim:2014uca}:
\begin{equation}
A^*=3-\exp \big[-k \frac{N_f}{N_c}+l\big] \ ,
\label{Kstar}
\end{equation}
where $k=16 T_R$ and $l=18.49-5.26 \ C_R /C_G$.

Performing a Taylor expansion of the integrand in Eq.\eqref{e7} and integrating term-by-term we can obtain the $n{\rm{th}}$-loop coefficients $b_{n,n-1}$ and check our criteria above for the existence of the safe fixed point~\cite{Shrock:2013cca,Dondi:2019ivp}. A preliminary investigation was performed in~\cite{Shrock:2013cca} up to 18th-loop order where it was also checked that the first 4-loops agree with the known perturbative results. It was found that, even though up to the 12th-loop order the resulting coefficients are scattered between the positive and negative values, starting from the 13th-loop order all $b_{n,n-1}$ are positive for the fundamental representation, two-index representations and symmetric or antisymmetric rank-3 tensors. Unfortunately, the positivity of the coefficients is insufficient to prove the stability of the series and determine its radius of convergence. The first complete study of the analytic properties of the leading nontrivial large-$N_f$ expansion appeared recently in~\cite{Dondi:2019ivp}. Here it was demonstrated that an analysis of the expansion coefficients to roughly 30 orders is required to establish the radius of convergence accurately, and to characterize the (logarithmic) nature of the first beta function singularity.

These studies agree with the existence of a singular structure at leading order in $1/N_f$ leading to a zero in the beta function. Although not a proof, see e.g. \cite{Alanne:2019vuk}, it can be viewed as lending support for the possible existence of an UV fixed point in these theories. These results have been confirmed when extended to theories with Yukawa interactions~\cite{Antipin:2018zdg,Kowalska:2017pkt,Alanne:2018ene} and employed to build realistic asymptotically safe extensions of the SM~\cite{Mann:2017wzh,Abel:2017rwl,Pelaggi:2017abg,Alanne:2018csn,Alanne:2019meg}.

Using the results above, we can sketch a complete phase diagram,
as a function of $N_c$ and $N_f$, for an $SU(N_c)$ gauge theory with fermionic matter in a given representation. A robust feature of this phase diagram is the line where  asymptotic freedom is lost, i.e.  $N_f^{AF}=11 C_G/(4T_R)$. As it is well known, decreasing $N_f$ slightly below this value  one achieves the perturbative Banks-Zaks infrared fixed point (IRFP), that at two loops yields $\alpha^*=-b_1/b_2$. This analysis has been extended to the maximum known order in perturbation theory in~\cite{Pica:2010xq,Ryttov:2010iz,Ryttov:2016ner}.

As the number of flavours decreases, the IRFP becomes strongly coupled and at some critical $N_f^{IRFP}$, is lost. The lower boundary of the conformal window has been estimated analytically in different ways~\cite{Appelquist:1986an,Miransky:1989nu,Sannino:2004qp,Dietrich:2006cm} and
summarised in~\cite{Sannino:2009za}. Combining these analytic results with nonperturbative lattice studies ~\cite{Leino:2017lpc, Leino:2017hgm,
Leino:2018qvq, Pica:2017gcb} defines the current state-of-the art.

Just above the loss of asymptotic freedom, as already mentioned,
Caswell~\cite{Caswell:1974gg} demonstrated that no perturbative UVFP can emerge. By continuity there should be a region in colour-flavour space where the resulting theory is nonabelian QED with an unavoidable Landau pole. This is the  {\it Unsafe QCD} region. The theories in this region are low energy effective field theories featuring a trivial IRFP. This means that one can expect the existence of a critical value of number of flavours $N_f^{Safe}$ above which safety emerges. This region extends to infinite values of $N_f$, i.e. the {\it Safe QCD} region~\cite{Antipin:2017ebo}.

\begin{figure*}[ht!]
\subfloat[Fundamental rep.]{\label{fig:CW1a} \includegraphics[width=0.45\textwidth]{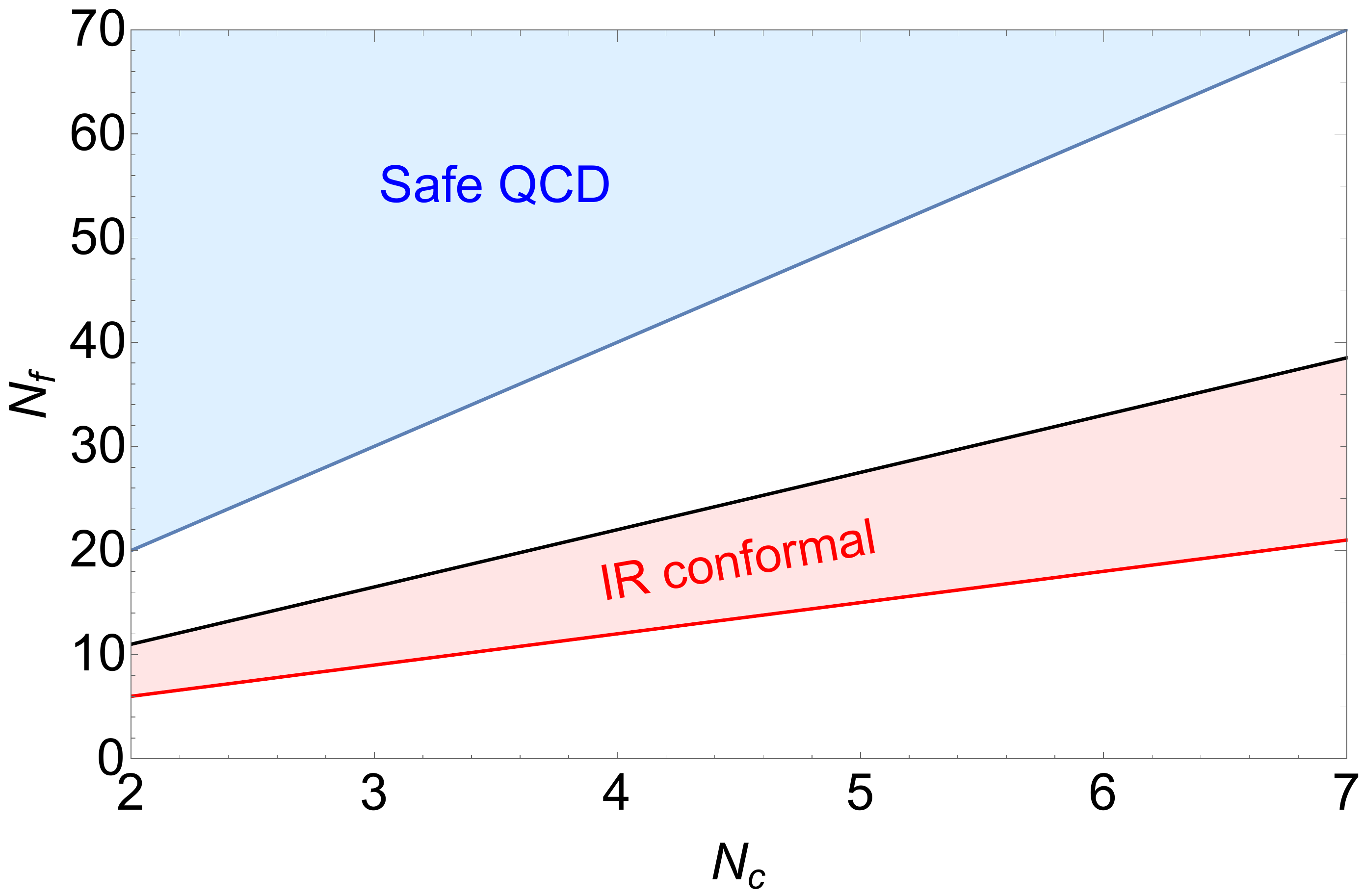}}
	\hfill
	\subfloat[Adjoint rep.]{\label{fig:CW1b} \includegraphics[width=0.45\textwidth]{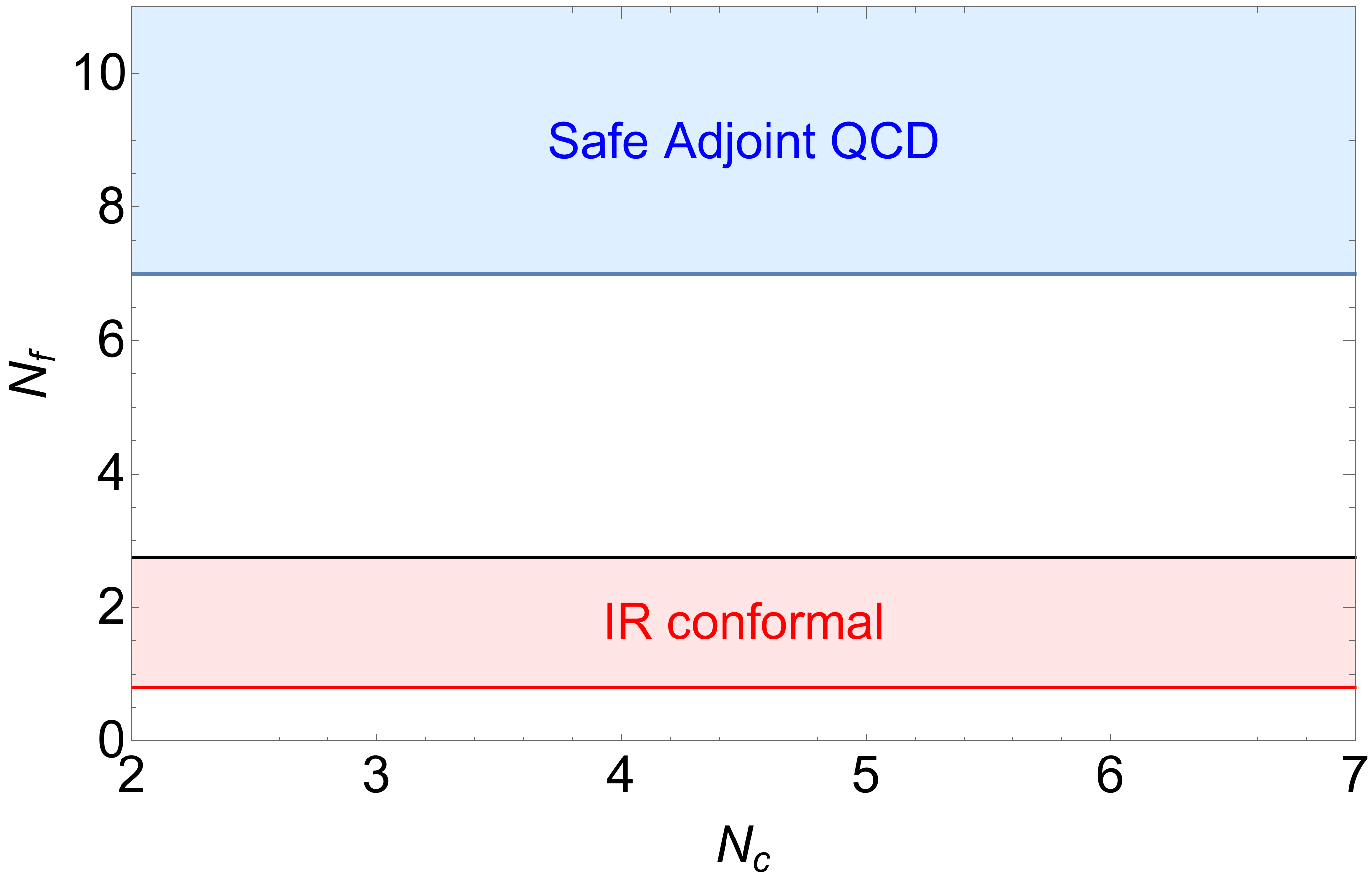}}
\caption{For ease of the reader we summarise here the phase diagram of $SU(N_c)$ gauge theories with fermionic matter in the fundamental (left-panel) and adjoint (right-panel) representation put forward in~\cite{Pica:2010xq,Antipin:2017ebo}. The shaded areas depict the corresponding conformal windows where the theories develop an IRFP (light red region) or an UVFP (light blue region).  The estimate of the lower boundary of the IRFP conformal window is taken from~\cite{Pica:2010xq}. }
\label{fig:CW}
\end{figure*}

For the fundamental representation, the leading $1/N_f$ expansion is applicable only for $N_c\lsim N_f/10$, while for the adjoint representation we find $N_f\gsim 7$ for any $N_c$. Following~\cite{Antipin:2017ebo}, it is sensible to use these values as first rough estimate of the lower boundary of the {\it Safe QCD} region. Altogether, these constraints allowed to draw the corresponding phase diagrams in~\cite{Antipin:2017ebo}.  For the reader's convenience we draw in Fig.~\ref{fig:CW} again the phase diagrams presented first in~\cite{Antipin:2017ebo} both for the fundamental (panel a) and adjoint representations (panel b).

Before specializing to the theories that we will investigate on the lattice let us comment also on the {\it safe} status of supersymmetric gauge theories. An UV safe fixed point can, in principle, flow to either a gaussian IR fixed point (non-interacting) or to an interacting IR fixed point. So far, for the non-supersymmetric case, we discussed the first class of theories because it is theoretically and phenomenologically important to assess whether non-asymptotically free theories can be UV complete, up to gravity. We provided an affirmative answer for gauge-Yukawa theories that are remarkably similar in structure to the Standard Model in~\cite{Litim:2014uca}. The situation for gauge-fermions is more involved and this is the reason we further investigate it here via first-principle lattice simulations.  Given the above the general conditions that must be (non-perturbatively) abided by non-asymptotically free supersymmetric theories to achieve safety  were put forward in~\cite{Intriligator:2015xxa}  generalising and correcting previous results of~\cite{Martin:2000cr}. To make the story short at least one chiral superfield must achieve a  large $R$ charge at the safe fixed point to ensure that the variation of the a-function between the safe and gaussian fixed points is positive as better elucidated in~\cite{Bajc:2016efj}. Models of this type were shown to exist in~\cite{Martin:2000cr,Bajc:2016efj,Bajc:2017xwx}. Another  possible way to elude the constraints~\cite{Intriligator:2015xxa,Bajc:2016efj} is to consider UV fixed points flowing to IR interacting fixed points. Within perturbation theory non-supersymmetric theories of this type were discovered in ~\cite{Esbensen:2015cjw}  and for supersymmetric theories in~\cite{Bond:2017suy,Bajc:2017xwx}.

Summarising the status for supersymmetric safety we can say that theories abiding the constraints of~\cite{Intriligator:2015xxa} exist. Nevertheless specialising~\cite{Intriligator:2015xxa} to the case in which asymptotic freedom is lost ($N_f \geq 3N_c$) for super QCD with(out)
a meson~\cite{Martin:2000cr,Intriligator:2015xxa} one can show that these theories are unsafe for any number of matter fields. This is in agreement with the  $1/N_f$ studies of the supersymmetric beta
functions~\cite{Ferreira:1997bi,Ferreira:1997bi,Martin:2000cr} as explained in~\cite{Dondi:2019ivp}. The unsafety of super QCD with(out) a meson should be contrasted with QCD at large number of flavors for which, as we argued above, safety is possible~\cite{Antipin:2017ebo} and QCD with a meson for which safety is a fact~\cite{Litim:2014uca}.

\begin{figure*}[ht!]
\centering
\begin{minipage}[t]{0.49\linewidth}
\centering
\includegraphics[width=0.8\linewidth]{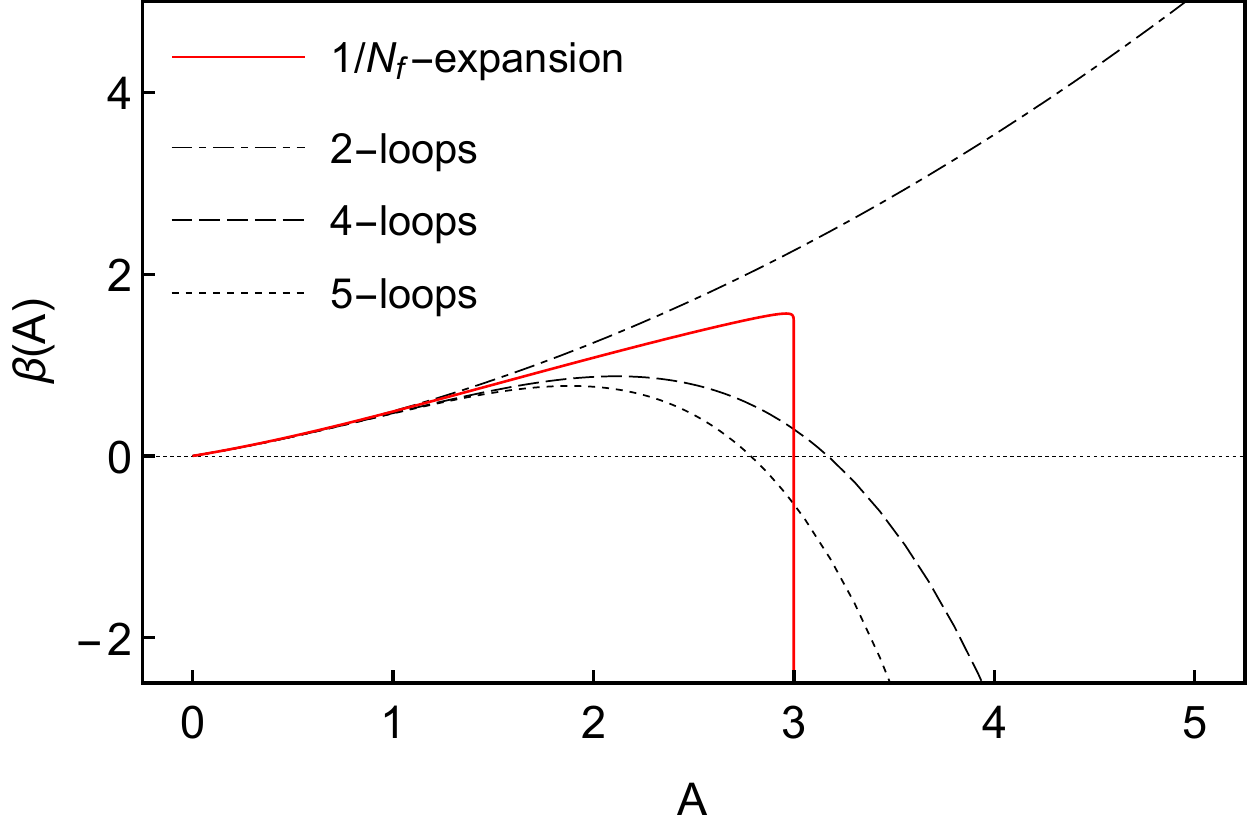}\\[8pt]
\includegraphics[width=0.8\linewidth]{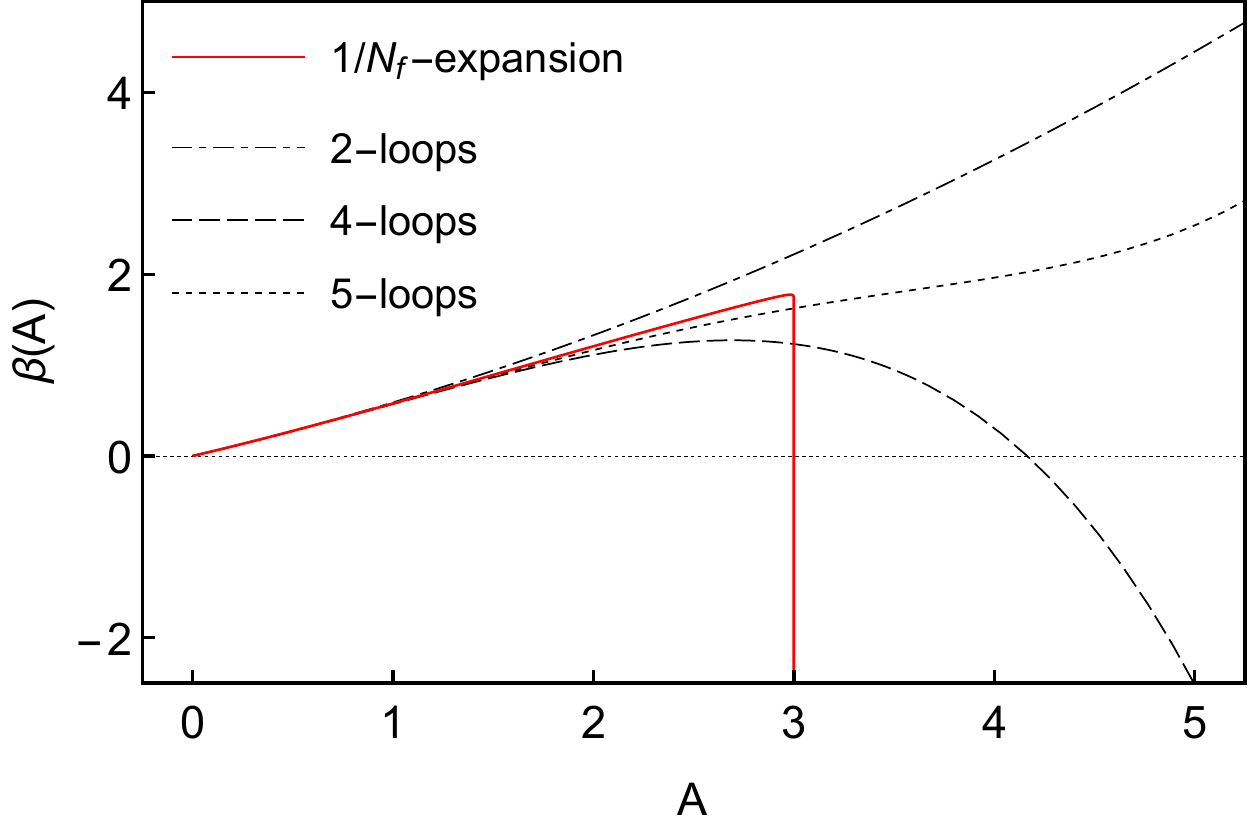}
\end{minipage}\hfill
\begin{minipage}[t]{0.49\linewidth}
\centering
\includegraphics[width=0.8\linewidth]{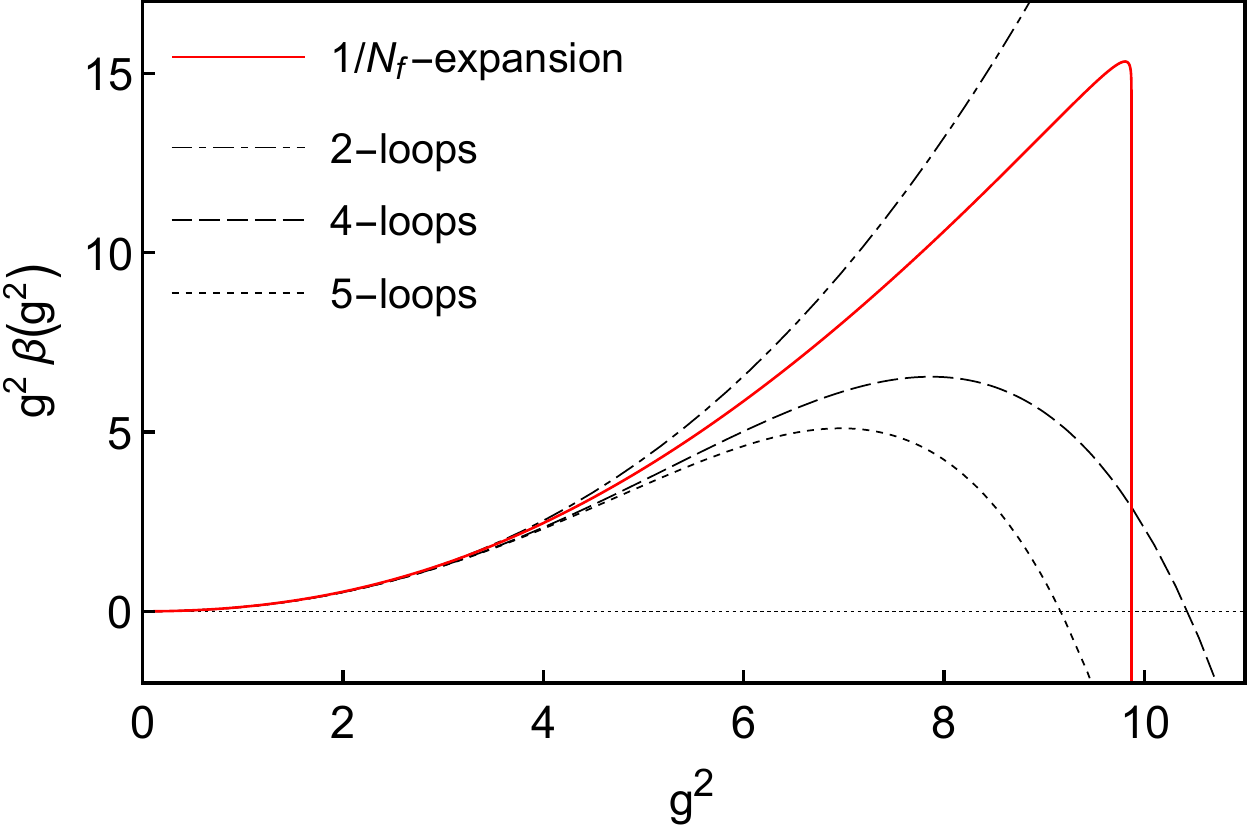}\\[8pt]
\includegraphics[width=0.8\linewidth]{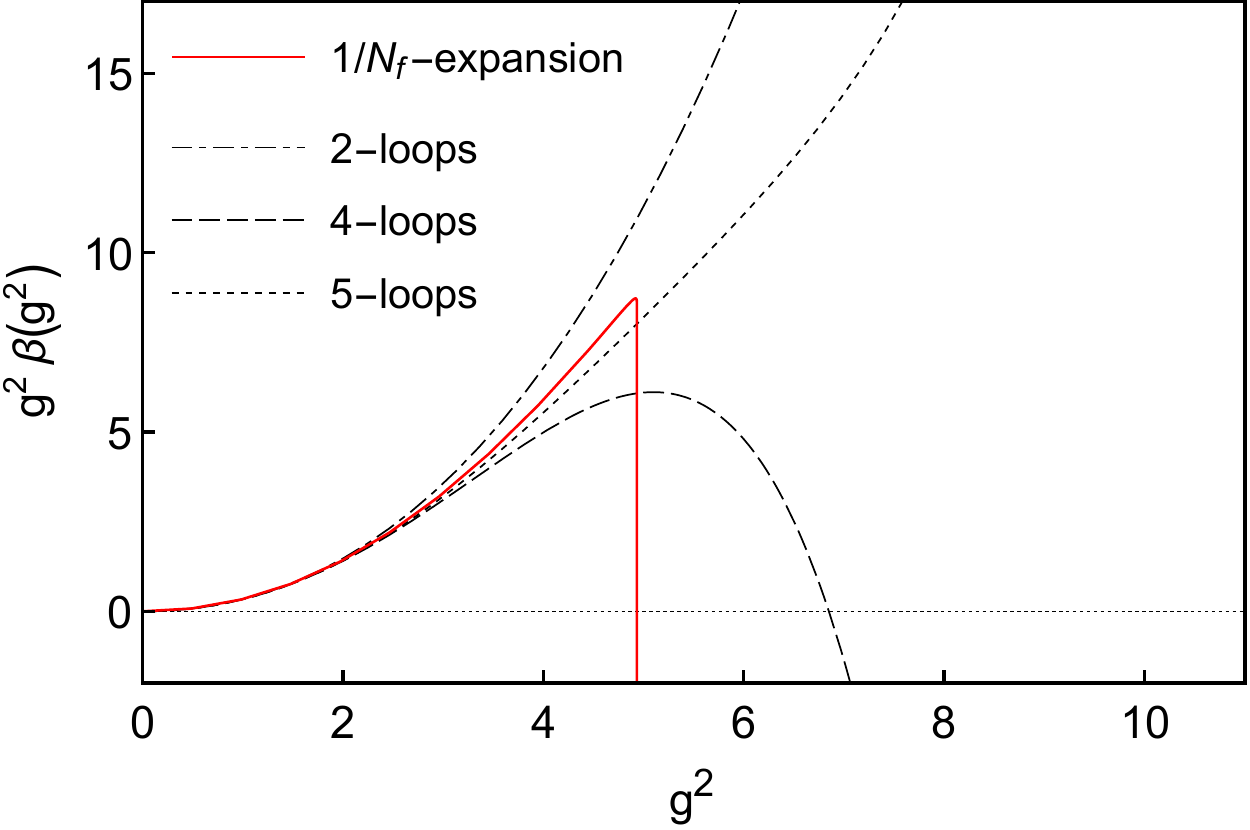}
\end{minipage}
\caption{Comparison of the leading order large-$N_f$ beta function (solid curve) and the perturbative beta function at two (dash-dotted curve), four (dashed curve) and five (dotted curve) loop orders. The upper panels are for $N_f=24$ and the lower ones for $N_f=48$. The left-hand panels show the beta function as defined in \eqref{e10} as function of $A=N_{f}\,T_{R}\,g^{2}/\ssof{4\,\pi^{2}}$, while the right-hand panels show the beta function as function of $g^{2}$ and rescaled with an additional factor of $g^{2}$.}
\label{fig:pertandLObetas}
\end{figure*}

After the supersymmetric parenthesis it is time to resume the investigation of the nonsupersymmetric conformal window 2.0. What has been reviewed so far clearly motivates  a nonperturbative study of the large number of flavour dynamics via lattice simulations.  Recalling that the rough estimate for the lower boundary  of the safe side of the conformal window 2.0   requires $N_f > 10 N_c$, it is clear that to minimise the number of flavours we should use $N_c = 2$ and with $N_f > 20$. Therefore in Fig.~\ref{fig:pertandLObetas} we present the leading $N_f$ beta function  for an SU(2) gauge theory with either $24$ or $48$ flavours. The large-$N_f$ beta function is shown by solid curves while the dotted curves show the
five-loop perturbative beta function. The upper panel is for the $N_f=24$ case and the
lower for the $N_f=48$ one. At leading order in  $1/N_f$ the beta functions support the presence of an UVFP for both flavours. It is instructive to show also the four and five loop perturbative results. These are the two theories we choose to investigate on the lattice, i.e. an SU(2) gauge theory with either 24 or 48 flavors transforming according to the fundamental representation of the underlying gauge group.

\section{Safety on the lattice}
\label{SoL}
\subsection{Lattice formulation}
\label{ssec:latform}
In this section we will define the model we consider and discuss the methods we use. Our treatment of the general features will be brief here as more detailed description can be found in~\cite{Rantaharju:2015yva,Leino:2017lpc}.
The model is defined by the lattice action
\begin{displaymath}
	S = S_G(U) + S_F(V) + c_{\rm SW}\delta S_{\rm SW}(V) \,,
        \label{lataction}
\end{displaymath}%
where $U$ is the standard SU(2) gauge link matrix, $V$ is smeared gauge link defined by hypercubic truncated stout smearing (HEX smearing)~\cite{Capitani:2006ni}, $S_G$ is the single plaquette gauge action and and $S_F$ is the Wilson fermion action with the clover term $S_{\rm SW}$.  We set the Sheikholeslami-Wohlert coefficient $c_{\rm SW} = 1$, which is a good approximation with HEX smeared fermions \cite{Rantaharju:2015yva}.


The coupling constant is measured using the gradient flow method with Dirichlet boundary conditions \cite{Ramos:2015dla}.  This method has been used successfully to measure the evolution of the coupling constant in SU(2) gauge theory with $N_f < 10$, motivating its use also here \cite{Rantaharju:2015yva,Leino:2017lpc}.
On a lattice of size~$L^4$ we use periodic boundary conditions at the spatial boundaries.
At the temporal boundaries $x_0=0$, $L$  we use Dirichlet boundary conditions by setting
the gauge link matrices to $U=V=1$ and the fermion fields to zero.
These boundary conditions enable simulations at vanishing fermion mass.

We run the simulations using the hybrid Monte Carlo algorithm
with 5th order Omelyan integrator~\cite{Omelyan:2003:SAI,Takaishi:2005tz}
and chronological initial values for the fermion matrix inversions~\cite{Brower:1995vx}.
The step length is tuned to have an acceptance rate of the order of 80\% or higher.
We run the simulations with bare couplings varying within the range
$\beta_L \equiv 4/g_0^2  \in[-1.,6.]$
and for each value of $\beta_L$ we tune the hopping parameter $\kappa$ to its critical value $\kappa_c(\beta_L)$, for which the absolute value of the PCAC fermion mass~\cite{Luscher:1996vw}
is of the order of $10^{-4}$ on a lattice of size $24^4$.
The obtained critical hopping parameter values, $\kappa_c(\beta_L)$, are then used for all used lattice sizes $L/a = 12$, $18$, $24$ and $30$.
A summary of the simulation parameters and corresponding PCAC quark masses, as well as the acceptance rates and accumulated statistics for each simulation, is given in the tables~\ref{tbl:nf24simparam} and \ref{tbl:nf48simparam}.

We note here that we include negative values for the inverse bare lattice gauge coupling $\beta_L$ in our study.  This is to compensate the effective positive $\beta_L$ shift induced by the Wilson fermions \cite{Hasenfratz:1993az,Blum:1994xb}.  This shift is proportional to the number of flavours and can therefore be substantial at large $N_f$.  Indeed, even at the smallest $\beta_L = -1$ at $N_f=48$ the lattice gauge field observables (for example, the plaquette) behave as if the effective gauge coupling remains positive.  Because we need to use strong coupling in the UV, we are forced to use small values for $\beta_L$.
A qualitatively similar effect has been observed with staggered fermions \cite{deForcrand:2012vh}.

To define the running coupling,
we apply the Yang-Mills gradient flow method~\cite{Narayanan:2006rf,Luscher:2009eq,Ramos:2015dla}.
This method defines a flow that smooths the gauge fields, removes UV divergences
and automatically renormalizes gauge invariant objects~\cite{Luscher:2011bx}.
The method is set up by introducing a fictitious flow time $t$
and studying the evolution of the flow gauge field $B_\mu(x,t)$ according to the flow equation
\begin{equation}
  \partial_t B_{\mu} = D_{\nu} G_{\nu\mu} \,,\;
  \label{eq:gradflow}
\end{equation}%
where $G_{\mu\nu}(x;t)$ is the field strength of the flow field $B_{\mu}$ and
$D_\mu=\partial_\mu+[B_\mu,\,\cdot\,]$.
The initial condition is $B_{\mu}(x;t=0) = A_\mu(x)$,
where $A_\mu$ is the original continuum gauge field.
In the lattice formulation the lattice link variable $U$
replaces the continuum flow field,
which is then evolved using 
the tree-level
improved L\"uscher-Weisz pure gauge action (LW)~\cite{Luscher:1984xn}.

\begin{table}[ht!]
\centering
\begin{tabular}{c | c | c | c | c | c | c | c | c | c | r}
 & &$m_{q}a$&\multicolumn{2}{|c|}{$L=12a$}&\multicolumn{2}{|c|}{$L=18a$}&\multicolumn{2}{|c|}{$L=24a$}&\multicolumn{2}{|c}{$L=30a$}\\
$\beta_{L}$ & $\kappa$ & {\small $[10^{-4}]$}& acc. & stat. & acc. & stat. & acc. & stat. & acc. & stat. \\\hline
 -0.3 & 0.131578 & \text{ 8.3(4)} & 0.88 & \text{5k} & 0.78 & \text{5k} & 0.83 & \text{4k} & 0.86 & \text{2.6k} \\
 -0.25 & 0.131354 & \text{ 9.5(5)} & 0.89 & \text{5k} & 0.78 & \text{5k} & 0.84 & \text{4k} & 0.88 & \text{2.6k} \\
 -0.2 & 0.131139 & \text{ 9.1(5)} & 0.89 & \text{5k} & 0.8 & \text{5k} & 0.84 & \text{4k} & 0.87 & \text{2.6k} \\
 -0.15 & 0.13093 & \text{ 8.1(5)} & 0.9 & \text{5k} & 0.8 & \text{5k} & 0.84 & \text{4k} & 0.88 & \text{2.6k} \\
 -0.1 & 0.130725 & \text{ 7.0(4)} & 0.9 & \text{5k} & 0.82 & \text{5k} & 0.86 & \text{4k} & 0.88 & \text{2.7k} \\
 -0.05 & 0.130519 & \text{ 5.3(6)} & 0.91 & \text{5k} & 0.82 & \text{5k} & 0.85 & \text{4k} & 0.9 & \text{2.6k} \\
 0.001 & 0.130322 & \text{ 0.1(4)} & 0.93 & \text{10k} & 0.84 & \text{4.9k} & 0.73 & \text{4k} & 0.89 & \text{2.7k} \\
 0.05 & 0.130129 & \text{ 1.2(4)} & 0.94 & \text{10k} & 0.84 & \text{4.8k} & 0.73 & \text{4.3k} & 0.9 & \text{2.3k} \\
 0.1 & 0.129944 & \text{-0.7(4)} & 0.94 & \text{10k} & 0.86 & \text{4.6k} & 0.74 & \text{4.2k} & 0.92 & \text{2.4k} \\
 0.15 & 0.129758 & \text{-0.3(4)} & 0.94 & \text{10k} & 0.86 & \text{4.5k} & 0.74 & \text{4.1k} & 0.91 & \text{2.7k} \\
 0.2 & 0.129579 & \text{-0.7(4)} & 0.94 & \text{10k} & 0.86 & \text{4.5k} & 0.77 & \text{3.8k} & 0.92 & \text{2.5k} \\
 0.25 & 0.129403 & \text{ 0.3(3)} & 0.94 & \text{10k} & 0.86 & \text{4.4k} & 0.78 & \text{4.1k} & 0.92 & \text{2.4k} \\
 0.3 & 0.129232 & \text{ 0.7(3)} & 0.95 & \text{10k} & 0.87 & \text{4.4k} & 0.78 & \text{4k} & 0.92 & \text{2.4k} \\
 1. & 0.127336 & \text{-0.1(3)} & 0.97 & \text{10k} & 0.94 & \text{4.4k} & 0.85 & \text{4k} & 0.96 & \text{2.4k} \\
 3. & 0.125525 & \text{-0.1(1)} & 0.99 & \text{10k} & 0.98 & \text{5k} & 0.98 & \text{4k} & 0.98 & \text{2k} \\
 4. & 0.125355 & \text{ 1.0(1)} & 0.99 & \text{10k} & 0.98 & \text{5k} & 0.96 & \text{4k} & 0.99 & \text{2.7k} \\
 6. & 0.125224 & \text{-0.2(1)} & 0.99 & \text{10k} & 0.98 & \text{5k} & 0.95 & \text{4k} & 0.99 & \text{3k} \\
\end{tabular}
\caption{The table shows for $N_{f}=24$ for each of our simulation points the value of $\beta_{L}$ and $\kappa$, as well as the resulting PCAC quark mass $m_{q}a$ (in units of $10^{-4}$), determined on the $L=24a$ lattice. Furthermore are shown for each system size, the acceptance rates of the HMC trajectories and the accumulated statistics, i.e. number of sampled configurations.}
\label{tbl:nf24simparam}
\end{table}

\begin{table}[ht!]
\centering
\begin{tabular}{c | c | c | c | c | c | c | c | c | c | r}
 & &$m_{q}a$&\multicolumn{2}{|c|}{$L=12a$}&\multicolumn{2}{|c|}{$L=18a$}&\multicolumn{2}{|c|}{$L=24a$}&\multicolumn{2}{|c}{$L=30a$}\\
$\beta_{L}$ & $\kappa$ &{\small $[10^{-4}]$}& acc. & stat. & acc. & stat. & acc. & stat. & acc. & stat. \\\hline
 -1. & 0.129534 & \text{ 2.2(6)} & 0.85 & \text{5k} & 0.82 & \text{5k} & 0.83 & \text{2.1k} & 0.78 & \text{1.5k} \\
 -0.5 & 0.128463 & \text{ 0.1(3)} & 0.88 & \text{5k} & 0.84 & \text{4.4k} & 0.85 & \text{1.9k} & 0.82 & \text{1.1k} \\
 -0.4 & 0.128269 & \text{ 0.5(3)} & 0.88 & \text{5k} & 0.86 & \text{4.9k} & 0.86 & \text{1.9k} & 0.83 & \text{1.1k} \\
 -0.3 & 0.128084 & \text{ 0.4(4)} & 0.9 & \text{4.6k} & 0.86 & \text{4.7k} & 0.86 & \text{1.8k} & 0.83 & \text{1.1k} \\
 -0.25 & 0.127996 & \text{-0.6(3)} & 0.9 & \text{4.1k} & 0.87 & \text{4.4k} & 0.86 & \text{1.8k} & 0.84 & \text{1.1k} \\
 -0.2 & 0.127905 & \text{ 0.1(3)} & 0.9 & \text{4.5k} & 0.87 & \text{4.1k} & 0.88 & \text{1.6k} & 0.87 & \text{1.3k} \\
 -0.15 & 0.12782 & \text{-0.1(3)} & 0.9 & \text{4.1k} & 0.88 & \text{4.4k} & 0.88 & \text{1.5k} & 0.86 & \text{1.2k} \\
 -0.1 & 0.127737 & \text{-0.4(3)} & 0.91 & \text{4.5k} & 0.89 & \text{4.8k} & 0.9 & \text{1.6k} & 0.84 & \text{1.2k} \\
 -0.05 & 0.127651 & \text{ 0.1(3)} & 0.91 & \text{4k} & 0.88 & \text{4.5k} & 0.89 & \text{1.6k} & 0.87 & \text{1.2k} \\
 0.001 & 0.12757 & \text{-0.6(2)} & 0.92 & \text{4k} & 0.88 & \text{4.8k} & 0.83 & \text{2.4k} & 0.88 & \text{1k} \\
 0.05 & 0.127487 & \text{ 1.8(3)} & 0.92 & \text{4k} & 0.89 & \text{5.1k} & 0.82 & \text{2.2k} & 0.9 & \text{1k} \\
 0.1 & 0.127413 & \text{ 0.4(2)} & 0.92 & \text{4k} & 0.89 & \text{4.6k} & 0.83 & \text{2.6k} & 0.87 & \text{1k} \\
 0.15 & 0.127344 & \text{-1.7(2)} & 0.93 & \text{4k} & 0.89 & \text{5.4k} & 0.82 & \text{2.3k} & 0.89 & \text{1k} \\
 0.2 & 0.127265 & \text{ 0.5(3)} & 0.94 & \text{4k} & 0.9 & \text{5k} & 0.83 & \text{2.1k} & 0.88 & \text{1k} \\
 0.25 & 0.127191 & \text{ 1.5(3)} & 0.94 & \text{4k} & 0.91 & \text{4.7k} & 0.85 & \text{2.4k} & 0.9 & \text{1k} \\
 0.3 & 0.127125 & \text{-0.1(2)} & 0.94 & \text{4k} & 0.91 & \text{4.4k} & 0.86 & \text{2.3k} & 0.9 & \text{1k} \\
 1. & 0.126332 & \text{ 0.5(2)} & 0.97 & \text{4k} & 0.95 & \text{5k} & 0.92 & \text{2k} & 0.95 & \text{1.1k} \\
 3. & 0.12544 & \text{-0.1(1)} & 0.98 & \text{2k} & 0.99 & \text{2.4k} & 0.97 & \text{2k} & 0.97 & \text{1.3k} \\
 6. & 0.125209 & \text{-0.3(1)} & 0.98 & \text{2k} & 0.98 & \text{2.3k} & 0.98 & \text{1.9k} & 0.98 & \text{1.3k} \\
\end{tabular}
\caption{The table shows for $N_{f}=48$ for each simulation point the value of $\beta_{L}$ and $\kappa$, as well as the resulting PCAC quark mass $m_{q}$ (in units of $10^{-4}$), determined on the $L=24a$ lattice. Furthermore are shown for each system size, the acceptance rates of the HMC trajectories and the accumulated statistics, i.e. number of sampled configurations.}
\label{tbl:nf48simparam}
\end{table}

%
%
%

\subsection{What to expect}
\label{ssec:whattoexpect}
The gradient flow equation has the form of a heat-equation 
which, in the case of a free gluon field in the Landau gauge, is solved by the heat kernel~\cite{Luscher:2010iy}
\[
H\of{y-x,t}\,=\,\frac{1}{\of{4\,\pi\,t}^{d/2}}\,\exp\of{-\frac{(y-x)^2}{4\,t}}\ .
\]
Hence, by evolving a gauge configuration with the gradient flow for some time $t$, one performs a
Gaussian smearing with smearing radius $\sigma=\tsqrt{2\,t}$, which means
that physical processes at length-scales below $2\,\sigma=\tsqrt{8\,t}=:\lambda\ssof{t}$ get effectively integrated out.
The scale $\mu\ssof{t}=1/\lambda\ssof{t}$ can therefore be interpreted as renormalization scale of gauge observables measured at flow time $t$.

The non-perturbative renormalized coupling at scale $\mu$~\cite{Luscher:2010iy} is then defined via energy measurement as:
\[
\gGF(\mu)\,=\,\mathcal{N}^{-1}t^2 \langle E(t) \rangle\vert_{t=1/\ssof{8\mu^2}}\ ,
\label{eq:gGF}
\]
where $\mathcal{N}$ is a normalization factor that has been calculated in~\cite{Fritzsch:2013je} to match the $\MSb$ coupling at tree level, and the gauge field energy density $E$ is measured only on the central time slice of the lattice: $x_0 = L/2$.

On the lattice, the finite lattice spacing $a$ (which is a function of the bare lattice parameters $\beta_L$ and $\kappa$) and the finite system size $L=N\,a$ (with $N$ being the number of lattice sites in each direction) restrict the renormalization scales, accessible with the gradient flow, to the range $1/L<\mu\ssof{t}< 1/a$.
Per construction, the flow always starts on the UV-side of this interval and evolves the gauge field towards the IR.
At the lattice scale $1/a$, the lattice theory deviates strongly from its continuum counterpart, and the gradient flow smearing scale $\lambda\ssof{t}=\tsqrt{8\,t}$ should therefore reach at least 2-3 lattice spacings before the renormalized coupling (or any other observable of the lattice gauge field) at flow time $t$ can be expected to behave like the corresponding continuum quantity.

\begin{figure}[ht]
\centering
  \includegraphics[width=0.44\textwidth]{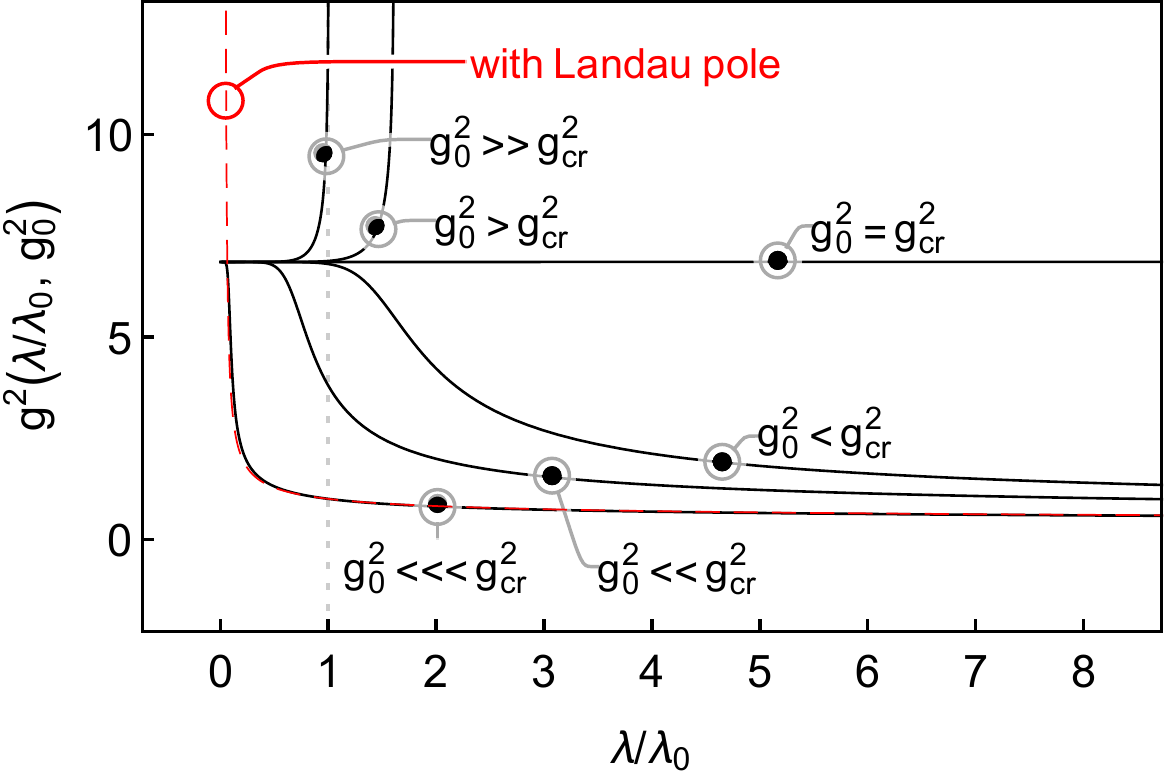}
\caption{A cartoon of the running coupling $g^{2}\ssof{\lambda/\lambda_{0},g_{0}^{2}}$ as a function of
increasing (relative) renormalization length scale $\lambda/\lambda_{0}$ in the presence of an interacting
UV fixed point, located at $g^{2}=g^{2}_{\rm cr}$, for different choices of the initial coupling $g_{0}^{2}=g^{2}\ssof{1,g_{0}^{2}}$
and corresponding reference scale $\lambda_{0}$.
Given the fixed point at $g^{2}_{\rm cr}$, the running coupling $g\ssof{\lambda/\lambda_{0},g^{2}_{0}}$
as a function of increasing $\lambda/\lambda_{0}\geq 1$ will
decrease if $g^{2}_{0}<g^{2}_{\rm cr}$, remain constant
if $g^{2}_{0}=g^{2}_{\rm cr}$ and increase if $g^{2}_{0}>g^{2}_{\rm cr}$.
The fixed point is repelling,
so that one can only flow away from it. If $g^{2}_{0}$ is much smaller than $g^{2}_{\rm cr}$,
the UVFP and Landau pole running couplings are almost indistinguishable for  $g^{2}\ssof{\lambda/\lambda_{0},g^{2}_{0}}<g^{2}_{\rm cr}$.}
\label{fig:gsqvsscaleuvfixed}
\end{figure}

One can now ask, how the lattice gradient flow coupling $\gGF\ssof{\lambda\ssof{t},\beta_{L}}$ should behave, as function of the flow scale $\lambda\ssof{t}$, if the theory possesses an interacting UV fixed point.
Before addressing this question for the lattice, let us recall how the running continuum coupling $g^{2}\ssof{\lambda/\lambda_{0},g^{2}_{0}}$, i.e.~the solution to the differential equation \eqref{e10} with $\mu=1/\lambda$, behaves as function of increasing $\lambda/\lambda_{0}>0$ for different choices of the initial condition or reference coupling $g^{2}_{0}=g\ssof{1,g^{2}_{0}}$ at reference scale $\lambda_{0}$, if there is an interacting UV fixed point at coupling $g^{2}=g^{2}_{\rm cr}$.
The behavior is illustrated schematically in Fig.~\ref{fig:gsqvsscaleuvfixed}: the fixed point is 
unstable and therefore, for increasing $\lambda/\lambda_{0}>1$,
the running coupling $g\ssof{\lambda/\lambda_{0},g^{2}_{0}}$ will:
\begin{itemize}
\item[(a)] decrease, if $g^{2}_{0}<g^{2}_{\rm cr}$,
\item[(b)] remain constant, if $g^{2}_{0}=g^{2}_{\rm cr}$ and
\item[(c)] increase, if $g^{2}_{0}>g^{2}_{\rm cr}$ .
\end{itemize}
We note that if $g^{2}_{0}$ is much smaller than $g^{2}_{\rm cr}$, the behavior of the running coupling
for $g^{2}\ssof{\lambda/\lambda_{0},g^{2}_{0}}<g^{2}_{\rm cr}$ can be almost indistinguishable from a Landau pole type behaviour.
The coupling constant evolution curves demonstrating an UV fixed point shown in
Fig.~\ref{fig:gsqvsscaleuvfixed} have been obtained by integrating the perturbative 4-loop beta function at $N_f=48$, which indeed has a zero at $g^2_{\rm cr} \approx 6.8$ (dashed curve in lower panels of Fig.~\ref{fig:pertandLObetas}).  The Landau pole curve has been obtained from the corresponding 1-loop
beta function.  While we naturally cannot expect purely perturbative results to be reliable, these curves give us qualitatively plausible scenarios for the coupling constant evolution.
%
We can conclude that it can be difficult to distinguish between UV fixed point and Landau pole behaviours by looking at the behavior of the running coupling for $g_{0}^{2}\ll g^{2}_{\rm cr}$.

Switching now to the the lattice, we can think of the lattice scale $a$ as defining the reference length scale $\lambda_{0}$ for the corresponding continuum theory. The gradient flow scale $\lambda\ssof{t}$, given in lattice units, then corresponds to the ratio of scales $\lambda/\lambda_{0}$ and we can for
$\lambda/\lambda_{0}\gg 2$ identify the running continuum coupling $g^{2}\ssof{\lambda/\lambda_{0},g_{0}^{2}}$ with the lattice gradient flow coupling
$\gGF\ssof{\lambda\of{t},\beta_{L}}$.
As already mentioned above, for $\lambda/\lambda_{0}\leq 2$ discretization effects are strong. Therefore
the lattice gradient flow coupling cannot be expected to
behave like the running coupling of the continuum theory\footnote{In fact, as we use in our simulations the clover Wilson fermion action with HEX smeared gauge links (meaning that the elementary cells on which the action is defined have a linear size of at least $2\,a$ instead of just $a$),
$\gGF\ssof{\lambda\of{t},\beta_{L}}$ and $g^{2}\ssof{\lambda/\lambda_{0},g^{2}_{0}}$ will even start to agree only for
$\lambda\ssof{t}=\lambda/\lambda_{0}\gtrsim 4$.}, which means that the relation between the bare inverse lattice coupling $\beta_{L}$ and the corresponding continuum coupling $g^{2}_{0}=g\ssof{\lambda/\lambda_{0}=1,g_{0}^{2}}$ cannot be read off from the value of $\gGF\ssof{\lambda\ssof{t},\beta_{L}}$ at flow scale $\lambda\ssof{t}=a$.

In order to verify the existence of an interacting UV fixed point via lattice simulations, one has to find a value of $\beta_{L}$, for which $a=\lambda_{0}$ is sufficiently small, so that the corresponding $g_{0}^{2}$ is bigger than $g_{\rm cr}^{2}$.
One should then observe, that the gradient flow coupling increases with increasing flow scale $\lambda\ssof{t}$, whereas for $g_{0}^{2}<g_{\rm cr}^{2}$, one will always flow towards the trivial IR fixed point.
For a theory which is IR- instead of UV-free, the dependency of the lattice spacing $a$ on the bare lattice parameter $\beta_{L}$ is unfortunately not necessarily such, that $a$ can become arbitrarily small, and it might therefore not be possible for $a=\lambda_{0}$ to reach values for which
$g_{0}^{2}>g^{2}_{\rm cr}$.

\begin{figure}[ht]
\centering
\includegraphics[width=0.43\textwidth]{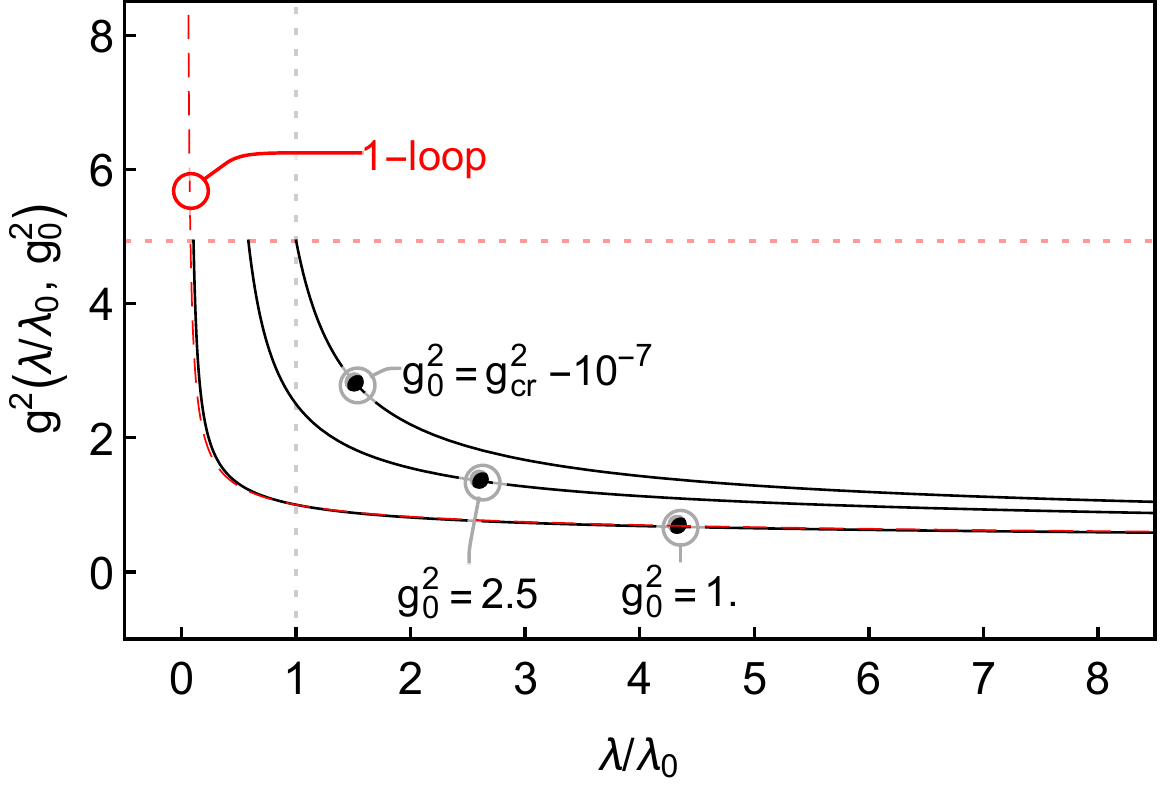}
\caption{The running coupling $g^{2}\ssof{\lambda/\lambda_{0},g_{0}^{2}}$ for the leading order  large-$N_{f}$ beta function from Eq.~\eqref{e8} with $N_{f}=48$ (c.f. lower panels of Fig.~\ref{fig:pertandLObetas}) as a function of increasing (relative) renormalizaion length-scale $\lambda/\lambda_{0}$ for different choices of initial coupling $g_{0}^{2}=g^{2}\ssof{1,g_{0}^{2}}$ and corresponding reference scale $\lambda_{0}$.
The red, horizontal dotted line indicates the location of the UV-fixed point, $g^{2}_{\rm cr}=4.9348022\ldots$.
In contrast to the four loop $N_f=48$ beta function used for the discussion in Fig.~\ref{fig:gsqvsscaleuvfixed},
which had a moderate slop at the fixed point, the slope of the leading order large-$N_{f}$ beta function at
the fixed point $g^{2}_{\rm cr}$ is very steep. This makes it almost impossible to draw the line of constant running coupling,
that would start at $g^{2}_{0}=g^{2}_{\rm cr}$ and remain there.
Consequently, the running coupling for $g^{2}_{0}<g^{2}_{\rm cr}$ looks even more as if there is a Landau pole at
$\lambda/\lambda_{0}\leq 1$, as can be seen by comparing for example the $g_{0}^{2}=1$ curve with the corresponding
1-loop result (dashed line), which has a Landau pole at $\lambda/\lambda_{0}=s_{\text{lp}}\approx 0.041$.
The leading order large-$N_{f}$ running coupling matches the 1-loop curve almost perfectly up to $g^{2}_{\rm cr}$,
where it suddenly stops. As this beta function ends right after the UV fixed point in a singularity at $-\infty$,
we do not have any information about the behavior of the running coupling above $g^{2}_{\rm cr}$.}
\label{fig:gsqvsscalelarenfexpfixed}
\end{figure}

The discovery of an interacting UV fixed point on the lattice in this way
is possible only if the true, non-perturbative beta function is such that the discussion in
connection with Fig.~\ref{fig:gsqvsscaleuvfixed} applies. As discussed above, the 4-loop beta  function was used to model the UV fixed point behaviour in Fig.~\ref{fig:pertandLObetas}.
The 4-loop beta function is smooth everywhere and its slope in the neighbourhood of the UV fixed point is moderate.
However, if the non-perturbative beta function behaves more like the leading order large-$N_{f}$ beta function (red curve in lower panels of Fig.~\ref{fig:pertandLObetas}), the situation is significantly more complicated, as the UV fixed point arises due to a singularity as
$\beta\to-\infty$.
This abrupt change makes the running coupling for $g^{2}_{0}<g^{2}_{\rm cr}$ look even more as if there is a Landau pole at
$\lambda/\lambda_{0}\leq 1$ (c.f. Fig.~\ref{fig:gsqvsscalelarenfexpfixed}), and because of the singularity right after the UV fixed point,
the behavior of the running coupling for $g^{2}_{0}>g^{2}_{\rm cr}$ is unknown.

\subsection{Results:  $N_f=24$  and $N_f=48$}
\begin{figure}[t]
\centering
\begin{minipage}[t]{0.49\linewidth}
\centering
\includegraphics[height=0.75\linewidth,keepaspectratio]{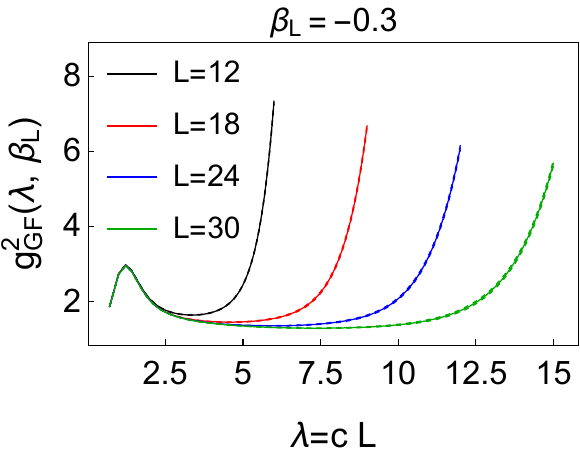}
\end{minipage}\hfill
\begin{minipage}[t]{0.49\linewidth}
\centering
\includegraphics[height=0.75\linewidth,keepaspectratio]{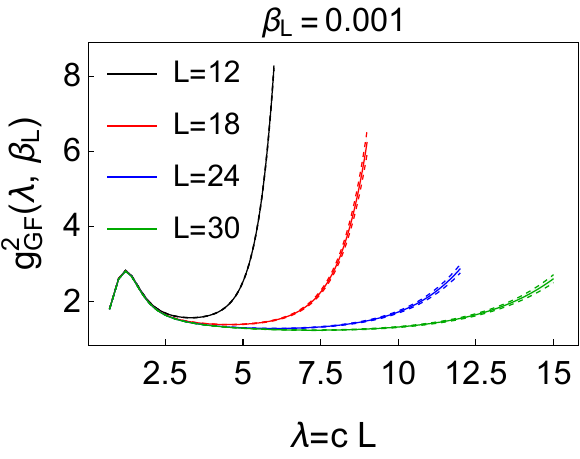}
\end{minipage}\\[3pt]
\begin{minipage}[t]{0.49\linewidth}
\centering
\includegraphics[height=0.75\linewidth,keepaspectratio]{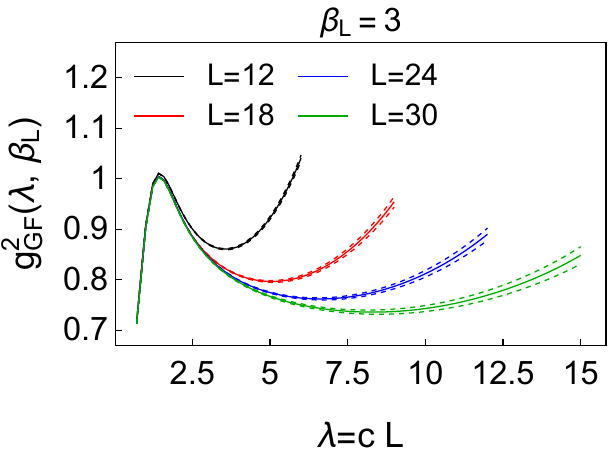}
\end{minipage}\hfill
\begin{minipage}[t]{0.49\linewidth}
\centering
\includegraphics[height=0.75\linewidth,keepaspectratio]{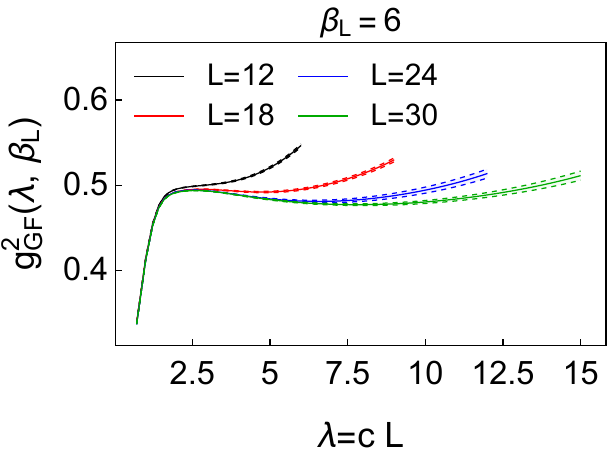}
\end{minipage}
\caption{The gradient flow couplings for $N_{f}=24$, measured at different values of the bare inverse lattice coupling $\beta_{L}$,  as functions of the flow scale $\lambda=\sqrt{8\,t}$ (where $t$ is the flow time in lattice units).}
\label{fig:ggfvslambdanf24}
\end{figure}

\begin{figure}[t]
\centering
\begin{minipage}[t]{0.49\linewidth}
\centering
\includegraphics[height=0.75\linewidth,keepaspectratio]{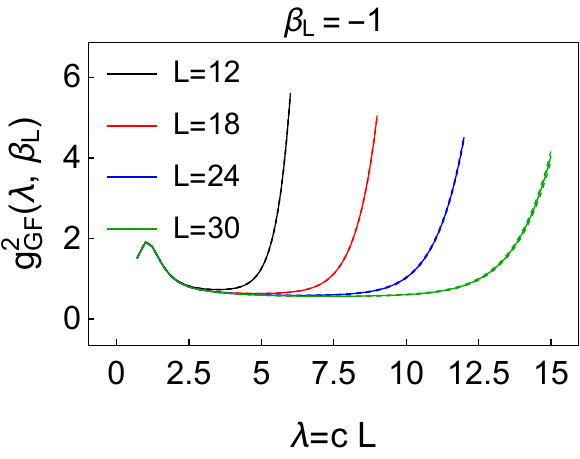}
\end{minipage}\hfill
\begin{minipage}[t]{0.49\linewidth}
\centering
\includegraphics[height=0.75\linewidth,keepaspectratio]{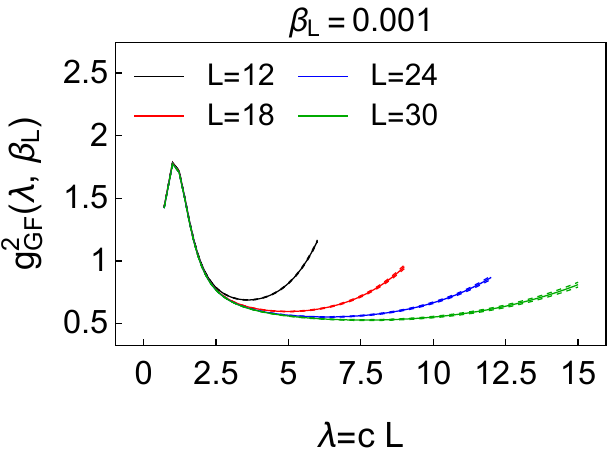}
\end{minipage}\\[3pt]
\begin{minipage}[t]{0.49\linewidth}
\centering
\includegraphics[height=0.75\linewidth,keepaspectratio]{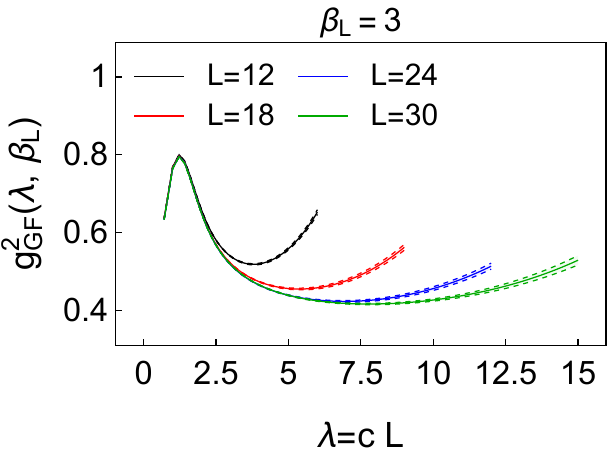}
\end{minipage}\hfill
\begin{minipage}[t]{0.49\linewidth}
\centering
\includegraphics[height=0.75\linewidth,keepaspectratio]{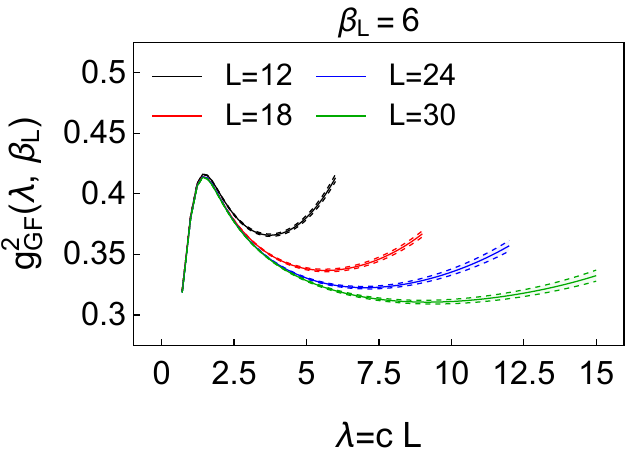}
\end{minipage}
\caption{The gradient flow couplings for $N_{f}=48$, measured at different values of the bare inverse lattice coupling $\beta_{L}$,  as functions of the flow scale $\lambda=\sqrt{8\,t}$ (where $t$ is the flow time in lattice units).}
\label{fig:ggfvslambdanf48}
\end{figure}

In Figs.~\ref{fig:ggfvslambdanf24} and \ref{fig:ggfvslambdanf48} we show examples of the  gradient flow running coupling $\gGF\ssof{\lambda,\beta_{L}}$ at $N_f=24$ and $48$, as functions of the gradient flow length-scale $\lambda=\tsqrt{8\,t}$.  The measurements are done on lattices of different sizes $L^4$, with $L/a=12,18,24,30$, and at different values of the bare inverse lattice coupling $\beta_{L}$.

These figures show that the measurements of $\gGF$ are dominated by lattice artefacts both at too small or at too large flow scales, $\lambda\lesssim 3a$ and $\lambda \gtrsim 0.3L$, respectively.  When the flow scale is of the order of lattice spacing or smaller large artefacts can be expected, and indeed, from the definition of the gradient flow coupling, Eq.~\eqref{eq:gGF}, we see that $\gGF \rightarrow 0$ as $\lambda \rightarrow 0$ ($t\rightarrow 0$).  This causes the characteristic ``peak'' at small $\lambda$ in Figs.~\ref{fig:ggfvslambdanf24} and \ref{fig:ggfvslambdanf48}, as $\gGF$ develops towards more continuum-like behaviour.  The use of the HEX smeared clover fermions also increases the range of interaction terms of the lattice action to $\gtrsim 3$-$4\,a$, which may also affect the flow at small $\lambda$.

Apart from these UV cut-off effects, the gradient flow is also affected by IR cut-off effects due to the finite lattice size.
In our case these seem to be particularly strong, as can be seen by comparing the curves for different system sizes in the individual panels of Figs.~\ref{fig:ggfvslambdanf24} and \ref{fig:ggfvslambdanf48}: the flow scales at which the $\gGF$ for different  system sizes start do deviate marks the scale at which IR cut-off effects start to dominate in the smaller of the two systems.
This seems here to happen already at scales around $\lambda\gtrsim 0.3L$ 
almost independently of the bare inverse lattice coupling $\beta_{L}$.


Due to the strong finite size and finite volume effects, the range of flow scales $\lambda$, for which the lattice gradient flow coupling
$\gGF\ssof{\lambda,\beta_{L}}$ can be expected to behave like the running coupling $g^{2}\ssof{\lambda/\lambda_{0},g^{2}_{0}}$ of the continuum theory, is approximately given by $\lambda\in\ssfof{3a,0.25L}$, where the lower bound of $3a$ is somewhat optimistic.  This renders the smallest lattices $L\leq 18$ of very limited use.

\begin{figure}[t]
\begin{tabular*}{\linewidth}[t]{p{0.05\linewidth}l}
  & \includegraphics[height=0.56\linewidth,keepaspectratio]{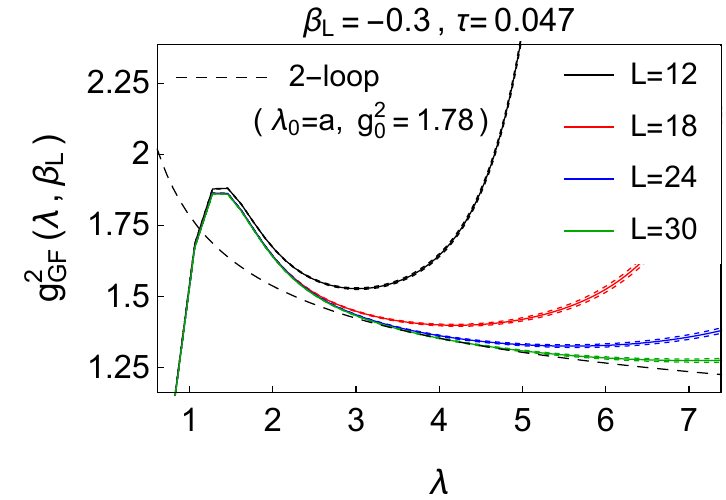}\\
  & \includegraphics[height=0.56\linewidth,keepaspectratio]{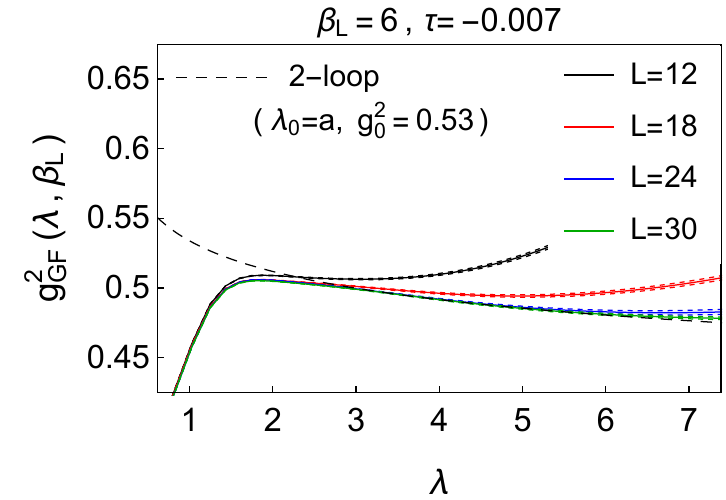}\\
\end{tabular*}
\caption{The gradient flow coupling for $N_{f}=24$ and bare inverse lattice coupling $\beta_{L}=-0.3$ (upper panel) and $\beta_{L}=6$ (lower panel)
as functions of the flow scale $\lambda$ after optimization of the parameter $\tau\of{\beta_{L}}$.
The value of $g^{2}_{0}=g\ssof{\lambda/\lambda_{0}=1,g^{2}_{0}}$, stated in the panel, can be used as a definition of an effective lattice-scale coupling for the lattice theory.}
\label{fig:impggfnf24}
\end{figure}

\begin{figure}[t]
\begin{tabular*}{\linewidth}[t]{p{0.05\linewidth}l}
  & \includegraphics[height=0.56\linewidth,keepaspectratio]{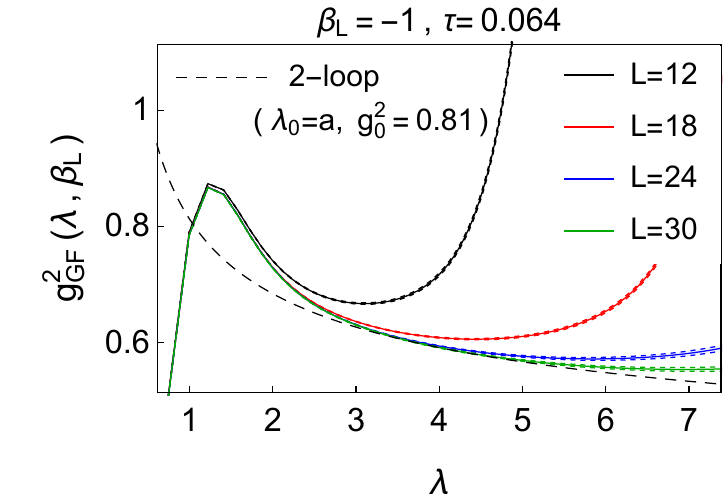}\\
  & \includegraphics[height=0.56\linewidth,keepaspectratio]{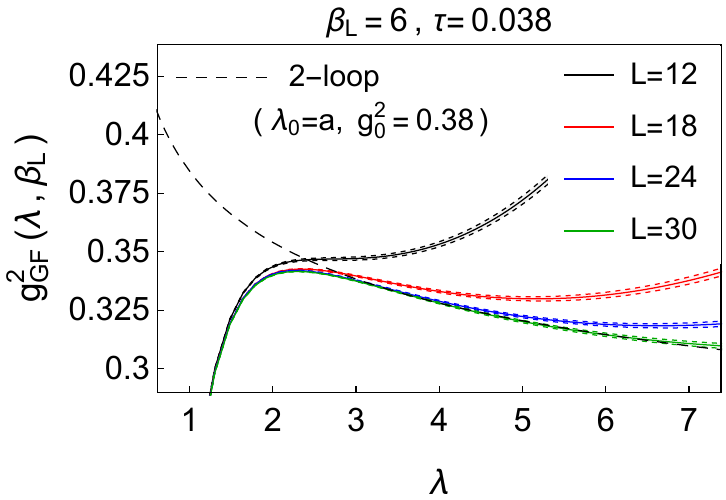}\\
\end{tabular*}
\caption{As in Fig.~\ref{fig:impggfnf24} but for $N_f=48$.}
\label{fig:impggfnf48}
\end{figure}

We can optimize the gradient flow coupling by adding a shift $\tau$ to the flow time \cite{Cheng:2014jba}:
\[
\gGF\ssof{\lambda,\beta_{L}}\,=\,\mathcal{N}^{-1}\,t^2\avof{E\ssof{t+\tau}}\big\vert_{t=\lambda^{2}/8}\ .\label{eq:imprggf}
\]
The effect of $\tau$ vanishes in the continuum limit (if it exists), and it can be tuned to get rid of most $\order\ssof{a^2}$ lattice artefacts.
Due to the very large finite size artefacts we cannot use the method used in \cite{Cheng:2014jba,Leino:2017lpc} to optimize $\tau$.
However, as it turns out that the gradient flow coupling obtained by our simulations are quite small, we can expect the scheme independent perturbative 2-loop beta function to be a very accurate approximation to the true beta function at large enough flow time (cf. Fig.~\ref{fig:pertandLObetas}).  For each $\beta_L$, we tune $\tau$ by matching the largest volume $L=30a$ gradient flow coupling to the 2-loop perturbative coupling over the interval $\lambda\in\ssfof{3a,5a}$.  We then use this value for $\tau$ for smaller volumes $L/a = 12,18,24$.
Examples for \eqref{eq:imprggf} at optimal $\tau\ssof{\beta_{L}}$ are shown in Fig.~\ref{fig:impggfnf24} ($N_{f}=24$) and Fig.~\ref{fig:impggfnf48} ($N_{f}=48$) for two different values of
$\beta_{L}$, together with the corresponding 2-loop running coupling.

\begin{figure}[t]
\centering\vspace{7pt}
\begin{minipage}[t]{0.499\linewidth}
\centering
{\small $N_{f}=24$}
\end{minipage}\hfill
\centering
\begin{minipage}[t]{0.499\linewidth}
\centering
{\small $N_{f}=48$}
\end{minipage}\\[3pt]
\begin{minipage}[t]{0.499\linewidth}
\centering
  \includegraphics[width=\textwidth]{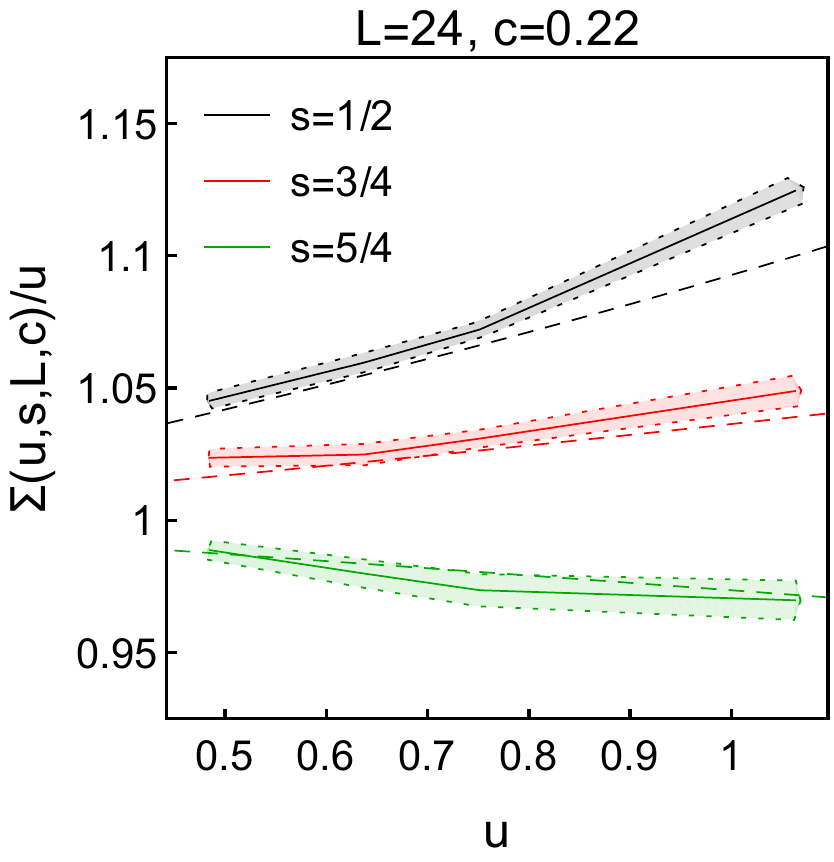}
\end{minipage}\hfill
\centering
\begin{minipage}[t]{0.499\linewidth}
\centering
  \includegraphics[width=\textwidth]{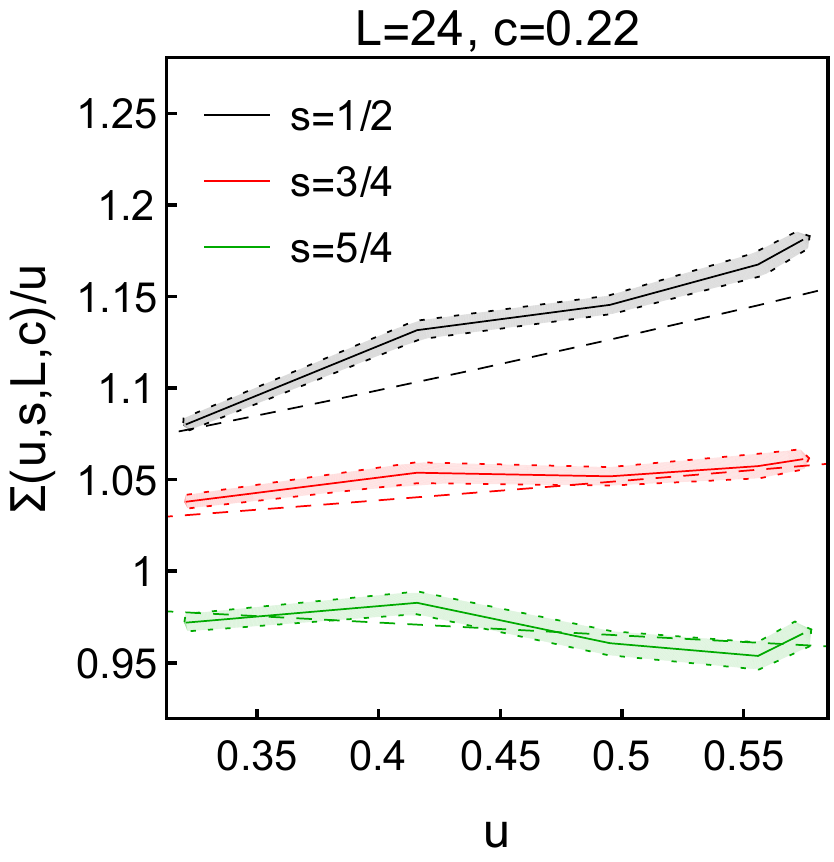}
\end{minipage}
\caption{$N_{f}=24$ (left) and $N_{f}=48$ (right) step scaling $\Sigma\of{u,s,L,c}/u$ as functions of $u$, with $\Sigma\ssof{u,s,L,c}=\gGF\ssof{s\,c\,L,\beta_{L}}\vert_{\beta_{L}:\gGF\ssof{c\,L,\beta_{L}}=u}$ defined in terms of the gradient flow coupling $\gGF\ssof{c\,L,\beta_{L}}$.
The latter can be interpreted as a non-perturbative definition of the renormalized coupling at scale $\lambda=c\,L\,a$.
For fixed $c$ and $\beta_{L}$, the ratio of the gradient flow couplings obtained with system sizes $L$ and $L'=s\,L$, i.e. $\gGF\ssof{c\,L',\beta_{L}}/\gGF\ssof{c\,L,\beta_{L}}$, can then be identified with the ratio of renormalized couplings $g^{2}\ssof{\lambda'}/g^{2}\ssof{\lambda}$ at scales $\lambda'$ and $\lambda$, where $\lambda'/\lambda=s$.
The dashed lines show the corresponding perturbative 2-loop results.}
\label{fig:stepscaling}
\end{figure}

In Fig.~\ref{fig:stepscaling} we show
the step scaling function
\[
\Sigma\ssof{u,s,L,c}=\gGFl{s L}\ssof{c\,s\,L,\beta_{L}}\big\vert_{\beta_{L}:\gGFl{L}\ssof{c\,L,\beta_{L}}=u} \,,
\label{eq:stepscaling}
\]
which tells us how the coupling constant evolves when the length scale where it is evaluated changes from $c\,L$ to $s\,c\,L$.  In the following we use $c=0.22$, i.e. the gradient flow
time is fixed so that $\sqrt{8t} = 0.22L$.  We are forced to use relatively small value for $c$ (in comparson with conventional $c=0.4 \ldots 0.5$, used in \cite{Leino:2017lpc}, for example) in order to avoid excessive finite volume effects.
The step scaling is particularly well suited for doing a 
continuum extrapolation (if it exists) of the running coupling.
For a conventional extrapolation, one would, however, need a set of lattice sizes $A=\cof{L_{1},\ldots,L_{n}}$, from which one can form several pairs $\ssof{L',L}$ that correspond to the same ratio $L'/L=s$.  The lattice volumes available to us do not allow this.  Furthermore, very large lattice artefacts at smaller volumes would make the extrapolation unreliable.

What we can do instead is to
compare the step scaling function for different $s$ directly with the corresponding 2-loop results in Fig.~\ref{fig:stepscaling}.
The agreement is reasonable, except for $s=1/2$, which involves the smallest system size $L=12$ for which the finite size and finite volume effects are particularly strong.

\begin{figure}[t]
\centering\vspace{7pt}
\begin{minipage}[t]{0.499\linewidth}
\centering
{\small $N_{f}=24$}
\end{minipage}\hfill
\centering
\begin{minipage}[t]{0.499\linewidth}
\centering
{\small $N_{f}=48$}
\end{minipage}\\[3pt]
\begin{minipage}[t]{0.499\linewidth}
\centering
  \includegraphics[width=\textwidth]{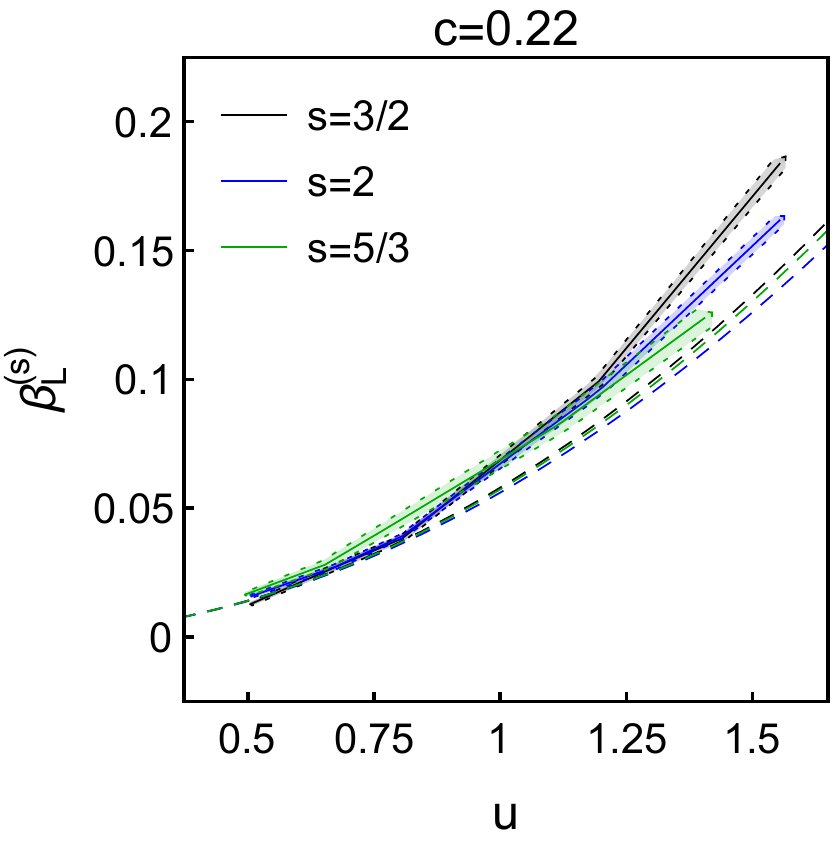}
\end{minipage}\hfill
\centering
\begin{minipage}[t]{0.499\linewidth}
\centering
  \includegraphics[width=\textwidth]{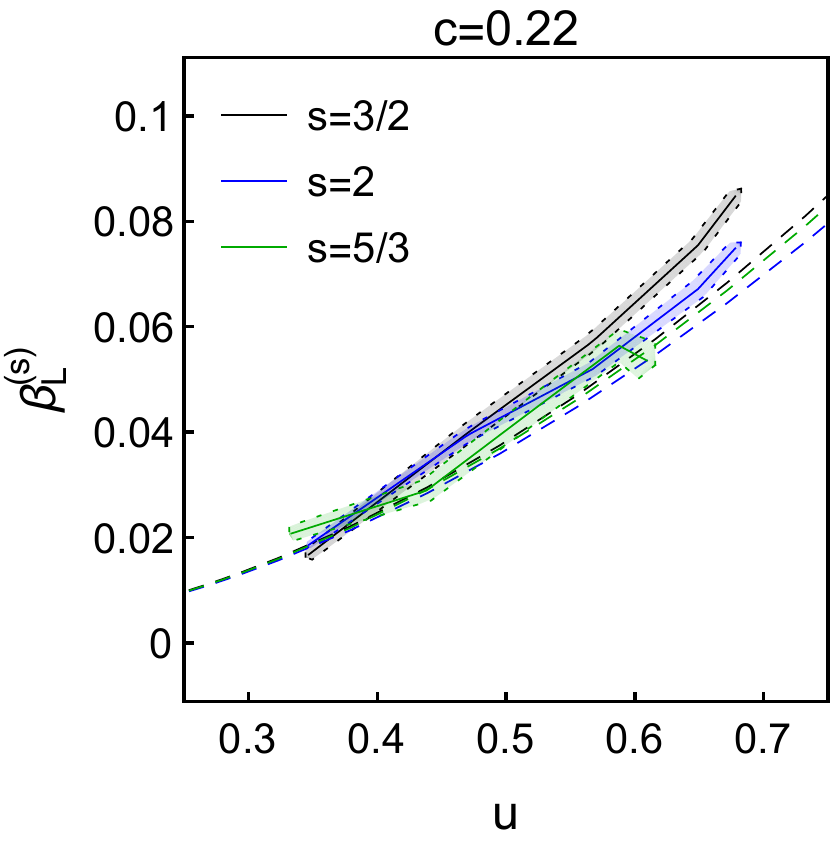}
\end{minipage}
\caption{$N_{f}=24$ (left) and $N_{f}=48$ (right) discrete beta function $\beta_{L}^{\of{s}}\of{u,L,c}=-\ssof{\gGFl{sL}\ssof{s\,c\,L,\beta_{L}}-\gGFl{L}\ssof{c\,L,\beta_{L}}}/\log\ssof{s^{2}}$ as functions of $u=\gGF\ssof{c\,L,\beta_{L}}$.
The dashed lines show the corresponding perturbative 2-loop results.}
\label{fig:discrbetafunc}
\end{figure}

A more transparent comparison of data from pairs of system size $\ssof{L',L}$ with different $s=L'/L$ can be obtained with the discrete beta function
\[
\beta_{L}^{\ssof{s}}\of{u,L,c}=-\frac{\gGFl{sL}\ssof{s\,c\,L,\beta_{L}}\,-\,u}{\log\ssof{s^{2}}}\bigg\vert_{\beta_{L}:\gGFl{L}\ssof{c\,L,\beta_{L}}=u}\ ,
\]
which approaches the conventional beta function in the limits $s\rightarrow 0$ or $g^2 \rightarrow 0$.
Because $\gGF$ is small, this is a good approximation of the true beta function.
The results are shown in Fig.~\ref{fig:discrbetafunc} for $c=0.22$ and $s=3/2=18/12$, $s=2=24/12$ and $s=30/18=5/3$.
For comparison, the corresponding results for the discrete beta function, evaluated with the 2-loop running coupling, are shown as well.  As we can observe, the lattice result approaches the 2-loop result as the volume increases.

From the results above we see that the measured gradient flow couplings are very small in the region where we can trust the measurements, at most $\gGF < 1.75$.  The couplings grow as the distance is reduced ($\lambda$ decreased), until lattice artefacts take over at $\lambda \lsim 3a$.  In order to reach strong couplings we use small values for the inverse bare coupling $\beta_L$ (even negative, as discussed in Sec.~\ref{ssec:latform}).  However, it turns out that if $\beta_L$ is small enough, the gradient flow coupling becomes almost independent of its value.  This behaviour is compatible with the Landau pole -like behaviour, as we will discuss in Sec.~\ref{sssec:effg2I} below.

\subsubsection{Effective $g^{2}_{0}$ I.}
\label{sssec:effg2I}

As we saw above, the gradient flow coupling makes sense only when the flow scale is large enough, $\gsim 3a$.  Nevertheless, it is of interest to relate $\gGF$ to some lattice ultraviolet scale effective coupling.
The inverse of $\beta_L = 4/g_0^2$ does not make much sense, because it can be negative.

Here we use the value of the plaquette (trace of the $1\times 1$ Wilson loop) as a proxy: it is an UV quantity on the lattice, and its expectation value is readily measurable.  To convert the plaquette values to effective couplings we use an ``inverse Monte Carlo'' procedure: we perform new simulations using a {\em pure gauge} SU(2) lattice theory with Wilson plaquette action (i.e. using only $S_G(U)$ in Eq.~\eqref{lataction}), and tune the pure gauge inverse coupling $\beta_L^{\rm PG}$ so that the plaquette expectation value matches that of the original theory.  The {\em effective bare coupling} is obtained by inversion
\begin{equation}
    g_{0,{\rm eff}}^2 = 4/\beta_L^{\rm PG}.
\end{equation}
This quantity is positive for all of our lattices.  Because the plaquette expectation value is, in practice, independent of the volume, $g_{0,{\rm eff}}^2$ is only a function of $\beta_L$ and $N_f$.  It increases monotonically as $\beta_L$ is decreased.  At our smallest $\beta_L$ values, $-0.3$ ($N_f=24$) and $-1.0$ ($N_f=48$), it reaches maximum values  $\sim 6$ and $\sim 5.3$, respectively.

\begin{figure}[t]
\centering\vspace{7pt}
\begin{minipage}[t]{0.499\linewidth}
\centering
{\small $N_{f}=24$}
\end{minipage}\hfill
\centering
\begin{minipage}[t]{0.499\linewidth}
\centering
{\small $N_{f}=48$}
\end{minipage}\\[3pt]
\begin{minipage}[t]{0.49\linewidth}
\centering
  \includegraphics[width=\textwidth]{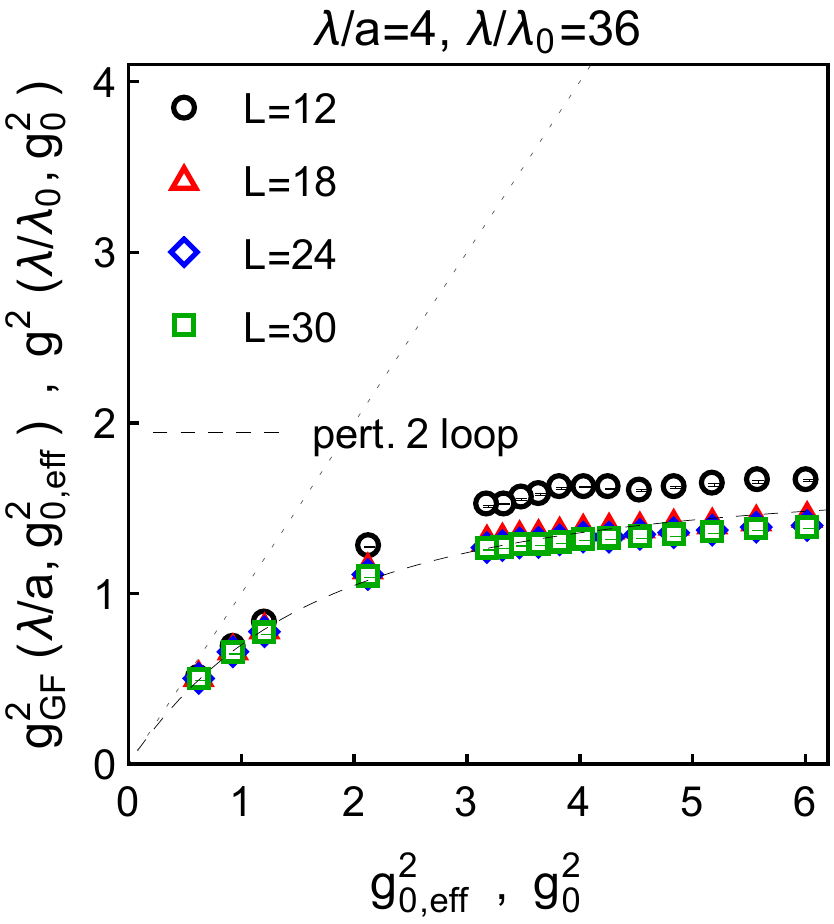}
\end{minipage}\hfill
\centering
\begin{minipage}[t]{0.49\linewidth}
\centering
  \includegraphics[width=\textwidth]{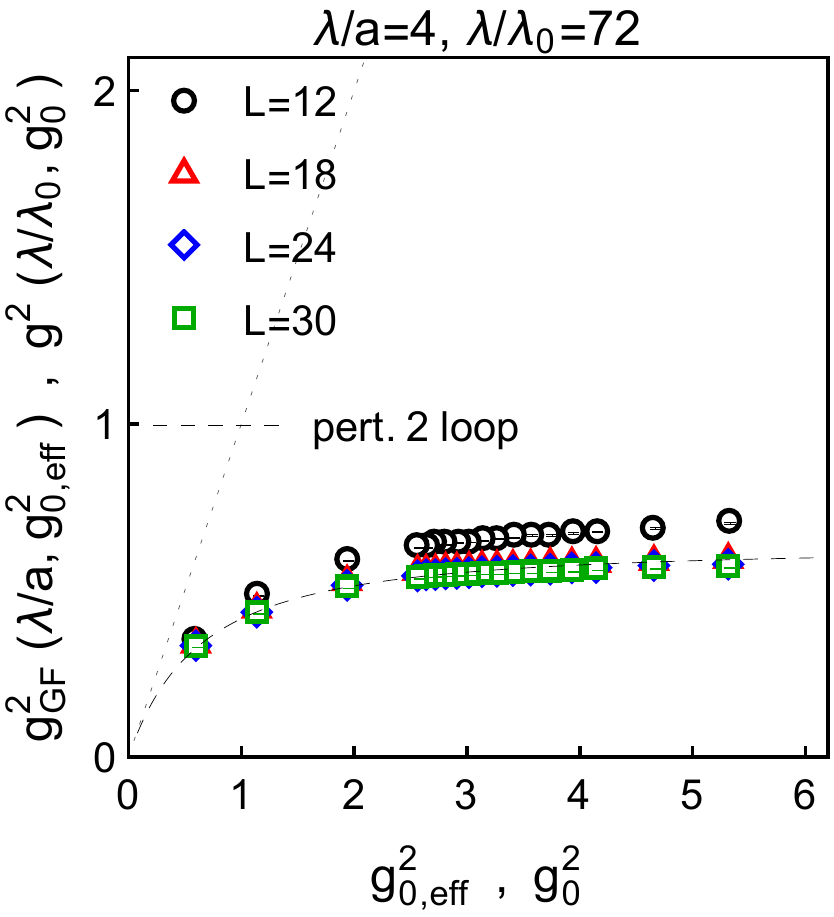}
\end{minipage}
\caption{The relation between the gradent flow coupling $\gGF\ssof{L,\beta_{L},c}$ at flow scale $\lambda=c\,L=4\,a$ and the effective bare coupling $\smash{g^{2}_{0,\text{eff}}\ssof{\beta_{L}}}$ at $N_{f}=24$ (left) and $N_{f}=48$ (right).
The dashed line is the perturbative 2-loop result, where the continuum couplings $g^{2}_{0}$ and $g^{2}\ssof{\lambda/\lambda_{0},g^{2}_{0}}$ are, respectively, identified with the $\smash{g^{2}_{0,\text{eff}}}$ and $\gGF\ssof{\lambda/a,\smash{g^{2}_{0,\text{eff}}}}$ from the lattice.
The $g_{0,{\rm eff}}^2 = \gGF$ line is shown with dots.}
\label{fig:gsqvsgsqeff0}
\end{figure}

In Fig.~\ref{fig:gsqvsgsqeff0} we plot the gradient flow couplings $\gGF$, measured at fixed flow scale $\lambda = 4a$, against $g_{0,{\rm eff}}$. The gradient flow coupling is almost independent of the volume, excluding $L/a = 12$, where flow scale $4a$ is large enough so that finite volume effects become important. We shall ignore $L/a=12$ in the following.

We observe the following behaviour: as $g_{0,{\rm eff}}^2 \rightarrow 0$, the ratio $\gGF/g_{0,{\rm eff}}^2  \rightarrow 1$, as dictated by universality at weak coupling.
On the other hand, as $g_{0,{\rm eff}}^2$ grows, $\gGF$ seems to approach a constant.  This is characteristic behaviour for an UV Landau pole: as the UV scale approaches the Landau pole, the UV scale coupling diverges.  Because $\gGF$ is evaluated at length scales which are by a constant factor larger than the UV length scale, the value of $\gGF$ will instead approach a constant value.

The couplings $g_{0,{\rm eff}}^2$ and $\gGF$ have been obtained using different schemas and thus their values cannot be compared without matching.  Nevertheless, we obtain a reasonable fit of
the data in Fig.~\ref{fig:gsqvsgsqeff0} with the 2-loop perturbative running coupling
$g^2(\lambda/\lambda_0, g_0^2)$ by
identifying $g^{2}_{0}\equiv g^{2}_{0,\text{eff}}$ and $g^{2}\ssof{\lambda/\lambda_{0},g^{2}_{0}}\equiv \gGF\ssof{\lambda/a,g^{2}_{0,\text{eff}}}$.
A good qualitative agreement is obtained
with the matching coefficients $\lambda_{0}=a/9$ for $N_{f}=24$ and $\lambda_{0}=a/18$ for $N_{f}=48$.
This implies that the effective lattice coupling $g_{0,{\rm eff}}^2$ corresponds to a reference scale, which is  about 9 resp. 18 times smaller than the lattice scale $a$.  The fit is compatible with the existence of the Landau pole, because the 2-loop beta function also features one.  However, we naturally cannot exclude the existence of an UVFP at stronger UV coupling than reached here.

The possible existence of the Landau pole gives a natural explanation for the small value of the gradient flow coupling: the lattice UV length scale is $\sim a$, whereas the gradient flow coupling $\gGF$ is evaluated at $\gsim 3a$.  Thus, in terms of energy, $\gGF$ is evaluated at scale $\mu_{\rm GF} < \mu_{\rm LP}/3$,
where $\mu_{\rm LP}$ is the scale where Landau pole appears.  This gives ample room for the coupling to decrease.  The actual value of the coupling depends on the details of the scheme.

\subsubsection{Effective $g^{2}_{0}$ II.}
\label{sssec:effg2II}
Another possibility to define an effective coupling for the lattice theory comes as a side product of the method we used to determine the optimal value of the $\tau$ parameter in the definition \eqref{eq:imprggf} of the improved gradient flow coupling, i.e.:
\[
\gGF\ssof{\lambda,\beta_{L};\tau}\,=\,\frac{t^2\avof{E\of{t+\tau}}}{\mathcal{N}\ssof{t+\tau}}\bigg\vert_{t=\lambda^{2}/8}\ .\label{eq:imprggf2}
\]
The procedure is as follows: from the lattice data for the gradient flow coupling $\gGF\ssof{\lambda,\beta_{L}}=\gGF\ssof{\lambda,\beta_{L};0}$ as function of flow scale $\lambda$, we produce data pairs $\ssof{\lambda_{i},\gGFl{i}}$, $i=1,2,\ldots$ , with $\gGFl{i}=\gGF\ssof{\lambda_{i},\beta_{L}}$ and $\lambda_{i}\in\ssfof{3,5}$.
Then, using that
\[
\gGF\ssof{\lambda,\beta_{L};\tau}\,=\,\frac{\lambda^{4}}{\ssof{\lambda^{2}+8\,\tau}^{2}}\gGF\ssof{\sqrt{\lambda^{2}+8\,\tau},\beta_{L}}\ ,
\]
we form the corresponding pairs for the $\tau$-shifted gradient flow coupling \eqref{eq:imprggf2}:
\[
\ssof{\lambda_{i}\ssof{\tau},\gGFl{i}\ssof{\tau}}\,=\,\sof{\sqrt{\vphantom{\lambda^{2}}\smash{\lambda_{i}^{2}}-8\,\tau},\frac{\ssof{\lambda_{i}^{2}-8\,\tau}^{2}}{\lambda_{i}^{4}}\,\gGFl{i}}\ .\label{eq:taushiftedlgsqpair}
\]
We then use the solution to the differential equation \eqref{e10} in the form
\[
\lambda_{\text{pt}}\ssof{g^{2},g^{2}_{0},\lambda_{0}}\,=\,\lambda_{0}\,\exp\bof{-\int_{g^{2}_{0}}^{g^{2}}\frac{\mathrm{d}u}{u\,\beta\ssof{u}}}\ ,\label{eq:pertlambdavsgsq}
\]
with $g^{2}\beta\ssof{g^{2}}$ being the perturbative 2-loop beta function, shown in Fig.~\ref{fig:pertandLObetas}, and minimize
\[
\chi^{2}\ssof{\tau}\,=\,\sum_{i}\sof{\lambda_{i}\ssof{\tau}-\lambda_{\text{pt}}\ssof{\gGFl{i}\ssof{\tau},g^{2}_{0}\ssof{\tau},\lambda_{0}\ssof{\tau}}}^{2}
\]
with respect to $\tau$, where $\ssof{\lambda_{0}\ssof{\tau},\gGFl{0}\ssof{\tau}}$ is given by the pair in \eqref{eq:taushiftedlgsqpair} for which the corresponding original $\lambda_{i}$, $i=1,2,\ldots$, is closest to the middle of the fitting interval $\lambda\in\ssfof{3,5}$.

After having determined in this way the optimal $\tau$, one also has fixed the values $\ssof{\lambda_{0},g^{2}_{0}}=\ssof{\lambda_{0}\ssof{\tau},\gGFl{0}\ssof{\tau}}$, for which \eqref{eq:pertlambdavsgsq} has within the given fit interval the best overlap with the $\tau$-shifted lattice data.
Solving now
\[
\lambda_{\text{pt}}\ssof{g^{2},g^{2}_{0},\lambda_{0}}=1
\]
for $g^{2}$, one obtains an effective coupling for the lattice theory at the cut-off scale $a$, which we call $g^{2}_{0,\text{eff2}}\ssof{\beta_{L}}$.

\begin{figure}[t]
\centering\vspace{7pt}
\begin{minipage}[t]{0.499\linewidth}
\centering
{\small $N_{f}=24$}
\end{minipage}\hfill
\centering
\begin{minipage}[t]{0.499\linewidth}
\centering
{\small $N_{f}=48$}
\end{minipage}\\[3pt]
\begin{minipage}[t]{0.49\linewidth}
\centering
  \includegraphics[width=\textwidth]{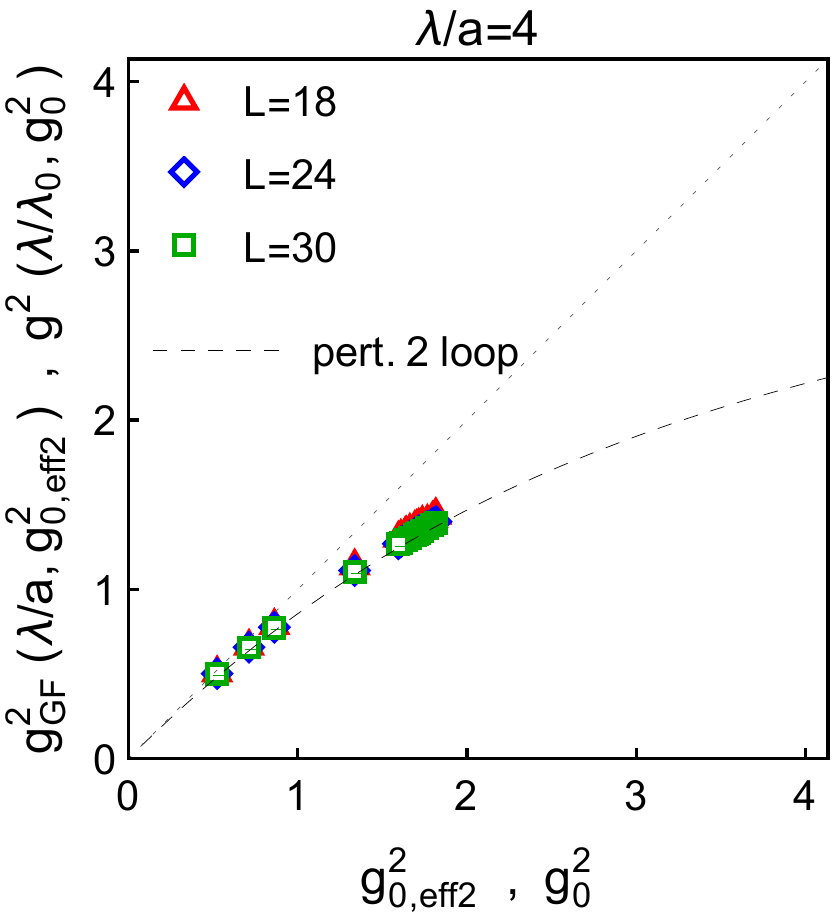}
\end{minipage}\hfill
\centering
\begin{minipage}[t]{0.49\linewidth}
\centering
  \includegraphics[width=\textwidth]{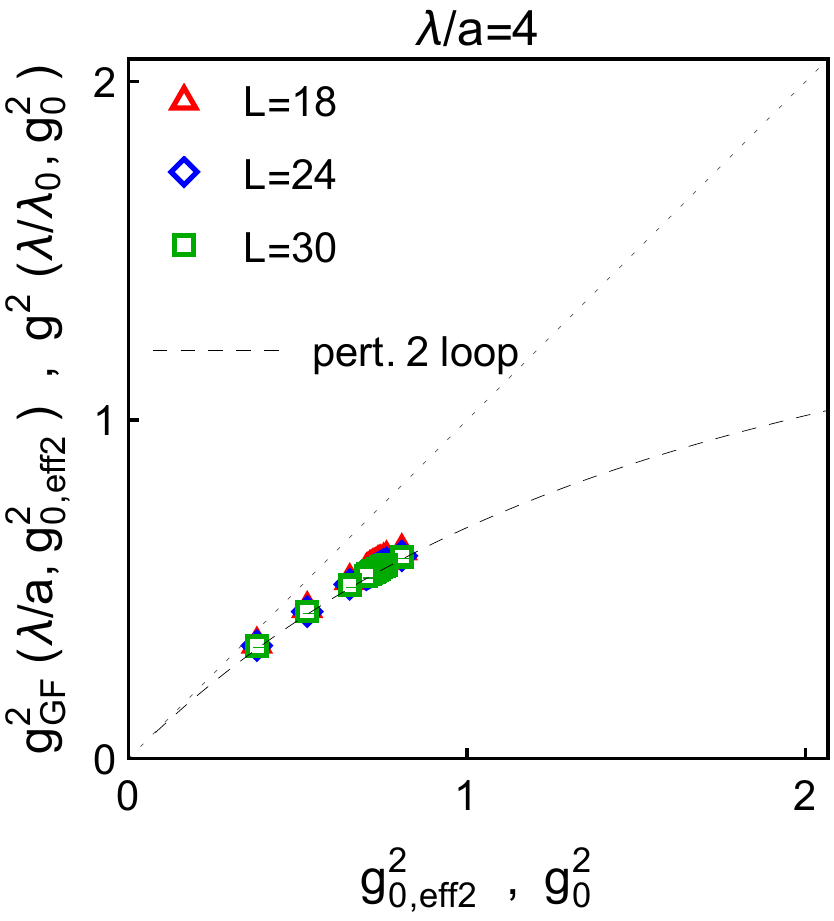}
\end{minipage}\\[3pt]
\begin{minipage}[t]{0.49\linewidth}
\centering
  \includegraphics[width=\textwidth]{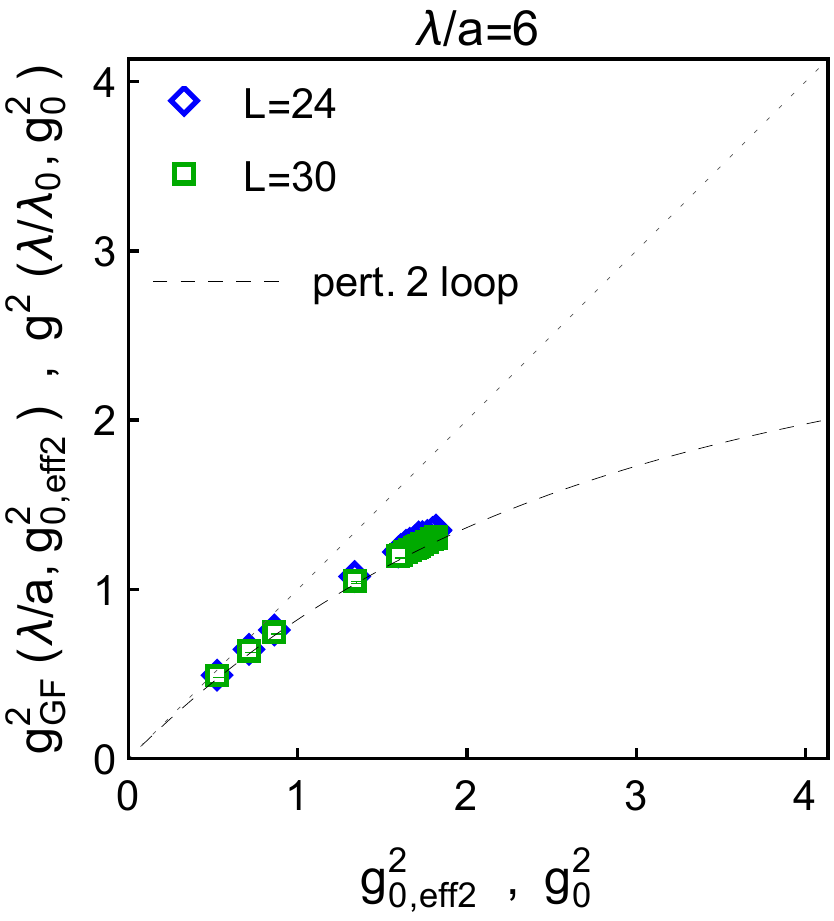}
\end{minipage}\hfill
\centering
\begin{minipage}[t]{0.49\linewidth}
\centering
  \includegraphics[width=\textwidth]{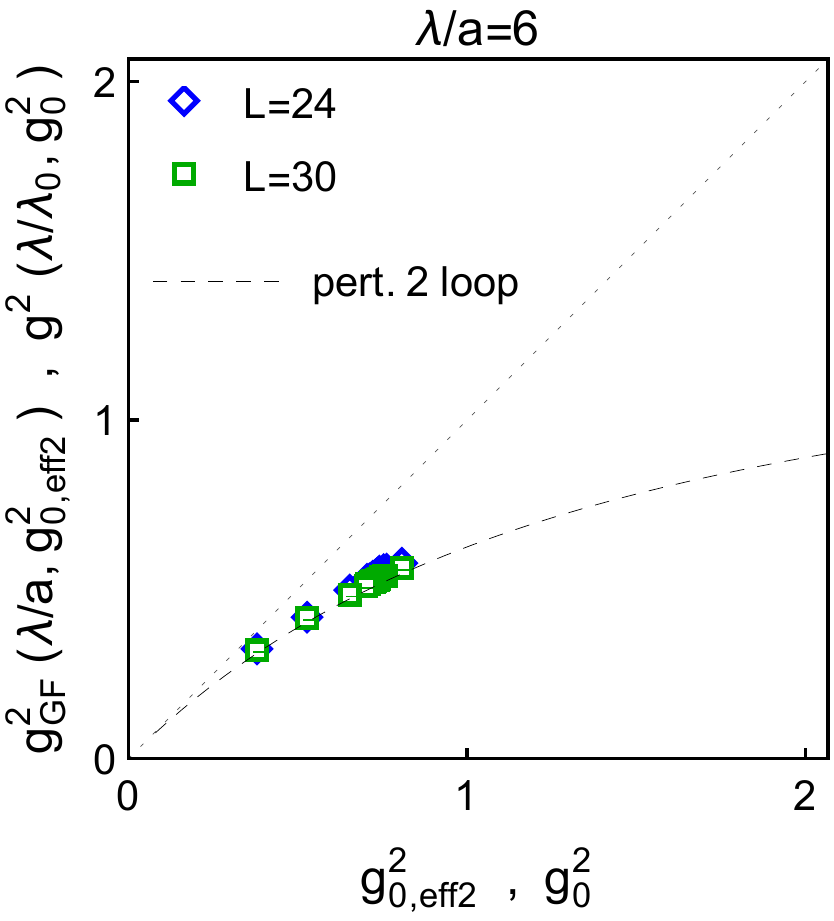}
\end{minipage}
\caption{The relation between $\gGF$ and effective lattice coupling $\smash{g^{2}_{0,\text{eff2}}\ssof{\beta_{L}}}$ at  $N_{f}=24$ (left) and $N_{f}=48$ (right), evaluated at
flow scale $\lambda=c\,L=4\,a$ (upper panels) and $6\,a$ (lower panels).  The dashed line is the corresponding perturbative 2-loop result, where the continuum couplings $g^{2}_{0}$ and $g^{2}\ssof{\lambda/\lambda_{0},g^{2}_{0}}$ are, respectively, identified with the $\smash{g^{2}_{0,\text{eff}}}$ and $\gGF\ssof{\lambda/a,\smash{g^{2}_{0,\text{eff2}}}}$ from the lattice.
The $g_{0,{\rm eff2}}^2 = \gGF$ line is shown with dots.}
\label{fig:gsqvsgsqeff02}
\end{figure}

From the derivation of the effective coupling $g^{2}_{0,\text{eff2}}\ssof{\beta_{L}}$, it is not a big surprise, that one finds in Fig.~\ref{fig:gsqvsgsqeff02} excellent agreement between lattice data and 2-loop result, when plotting the gradient flow coupling at fixed flow scale $\lambda$ (in lattice units) as function of $g^{2}_{0,\text{eff2}}\ssof{\beta_{L}}$, identifying the perturbative $g^{2}\ssof{\lambda/\lambda_{0},g^{2}_{0}}$ and $\gGF\ssof{\lambda,\beta_{L}}$, with $\lambda_{0}\equiv a$ and $g^{2}_{0}\equiv g^{2}_{0,\text{eff2}}\ssof{\beta_{L}}$.

\section{Conclusions}
\label{Conc}
The UV behavior of gauge-fermion theories as a function of the number of matter fields poses a fundamental problem on our understanding of quantum field theory, both perturbatively and nonperturbatively. Recent theoretical developments suggest that at inifinite number of flavors an interacting UV fixed point could exist making these theories asymptotically safe.

In this paper we have discussed in detail the categorization of gauge-fermion theories based on their possible UV behaviors. Then we described our computational setting to investigate the renormalization group evolution of SU(2) gauge theory with 24 or 48 Dirac fermions. Finally, we presented the results of our first pioneering analysis.

We found that with the methodologies developed in our past work, the non perturbative computations can be successfully carried out. We have demonstrated that our results match well with perturbation theory. On the other hand, we were unable to reach lattices where strong renormalized couplings could be controllably obtained.
We used the gradient flow procedure to determine the renormalized coupling on the lattice.  By construction gradient flow coupling can be evaluated at length scales which are at least a few lattice spacings.  This leaves room for a Landau pole or otherwise large couplings to exist at shorter distances (in the lattice ultraviolet limit).  We observed indications of this behaviour by defining an effective lattice ultraviolet coupling.  The observed behaviour is compatible with the existence of a Landau pole, but we cannot exclude an ultraviolet fixed point at even stronger couplings than reached here.
 Consequently, the true continuum behavior of these theories in the deep ultraviolet remain undetermined.

Nevertheless, our study provides an important milestone towards establishing the ultaviolet fate of gauge-fermion theories when asymptotic freedom is lost.  The results
further motivate investigations of gauge theories at very large number of fermion fields.

\section{Acknowledgement}
\label{Aknwldg}
The support of the Academy of Finland grants 308791, 310130 and 320123 is acknowledged. The authors wish to acknowledge CSC - IT Center for Science, Finland, for computational resources.
V.L is supported by the DFG cluster of excellence Origins. The work of F.S. is partially supported by the Danish National Research Foundation under the grant DNRF:90.

\end{document}